\documentclass[useAMS,usenatbib,nofootinbib]{mn2e}

\usepackage[bookmarks,bookmarksopen,colorlinks,linkcolor={blue},citecolor={green},urlcolor={red}]{hyperref}

\usepackage{amsmath}
\usepackage{url}
\usepackage{natbib}
\usepackage{rotating}

\title[SCUBA-2: The 10000 pixel bolometer camera on the JCMT]
{SCUBA-2: The 10000 pixel bolometer camera on the James Clerk Maxwell Telescope}

\author[W.~S.~Holland~et~al.]{
  \parbox[t]{\textwidth}{
    W.~S.~Holland$^{1,2}$\thanks{E-mail:~wayne.holland@stfc.ac.uk},
    D.~Bintley$^{3}$,
    E.~L.~Chapin$^{3,4}$\thanks{Present address: XMM SOC, ESAC, Apartado 79, 28691 Villaneueva de la Canada, Madrid, Spain},
    A.~Chrysostomou$^{3}$\thanks{Present address: School of Physics, Astronomy and Mathematics, University of Hertfordshire, Hatfield, 
Hertfordshire AL10 9AB},
    G.~R.~Davis$^{3}$,
    J.~T.~Dempsey$^{3}$,
    W.~D.~Duncan$^{1,5}$\thanks{Present address: Intellectual Ventures, 3150 139th Ave SE, Building 4, Bellevue, WA 98005, USA},
    M.~Fich$^{6}$,
    P.~Friberg$^{3}$,
    M.~Halpern$^{4}$,
    K.~D.~Irwin$^{5}$,
    T.~Jenness$^{3}$,
    B.~D.~Kelly$^{1}$,
    M.~J.~MacIntosh$^{1}$,
    E.~I.~Robson$^{1}$,
    D.~Scott$^{4}$,
    P.~A.~R.~Ade$^{7}$,
    E.~Atad-Ettedgui$^{1}$,
    D.~S.~Berry$^{3}$,
    S.~C.~Craig$^{3}$\thanks{Present address: National Solar Observatory, Advanced Technology Solar Telescope, 950 N. Cherry 
Avenue, Tucson AZ 85719, USA}, 
    X.~Gao$^{1}$,
    A.~G.~Gibb$^{4}$,
    G.~C.~Hilton$^{5}$,
    M.~I.~Hollister$^{2}$\thanks{Present address: California Institute of Technology, 1200 East California Boulevard, 
Pasadena, CA 91125, USA},
    J.~B.~Kycia$^{6}$,
    D.~W.~Lunney$^{1}$,
    H.~McGregor$^{1}$\thanks{Present address: Institute for Astronomy, 2680 Woodlawn Drive, Honolulu, HI 96822, USA},
    D.~Montgomery$^{1}$,
    W.~Parkes$^{8}$,
    R.~P.~J.~Tilanus$^{3}$,
    J.~N.~Ullom$^{5}$,
    C.~A.~Walther$^{3}$,
    A.~J.~Walton$^{8}$,
    A.~L.~Woodcraft$^{7}$\thanks{Present address: QMC Instruments, School of Physics \& Astronomy, Cardiff University, 5 
The Parade, Cardiff CF24 3AA},
    M.~Amiri$^{4}$,
    D.~Atkinson$^{1}$,
    B.~Burger$^{4}$,
    T.~Chuter$^{3}$,
    I.~M.~Coulson$^{3}$,
    W.~B.~Doriese$^{5}$,
    C.~Dunare$^{8}$,
    F.~Economou$^{3}$\thanks{Present address: National Optical Astronomy Observatory, 950 N. Cherry Avenue, Tucson, AZ 85719, USA},
    M.~D.~Niemack$^{5}$,
    H.~A.~L.~Parsons$^{3}$,
    C.~D.~Reintsema$^{5}$,
    B.~Sibthorpe$^{1}$,
    I.~Smail$^{9}$,
    R.~Sudiwala$^{7}$,
    H.~S.~Thomas$^{3}$
  }
  \\
  \\
  $^{1}$UK Astronomy Technology Centre, Royal Observatory, Blackford Hill, Edinburgh EH9 3HJ\\
  $^{2}$Institute for Astronomy, University of Edinburgh, Royal Observatory, Blackford Hill Edinburgh, EH9 3HJ\\
  $^{3}$JointAstronomy Centre, 660 N. A`oh\={o}k\={u} Place, University Park, Hilo, Hawaii 96720, USA\\
  $^{4}$Department of Physics \& Astronomy, University of British Columbia, 6224 Agricultural Road, Vancouver BC V6T 1Z1, Canada\\
  $^{5}$National Institute of Standards and Technology, 325 Broadway, Boulder CO 80305, United States\\
  $^{6}$Department of Physics, University of Waterloo, Waterloo, Ontario N2L 3G1, Canada\\
  $^{7}$School of Physics \& Astronomy, Cardiff University, 5 The Parade, Cardiff CF24 3AA\\
  $^{8}$Scottish Microelectronics Centre, University of Edinburgh, West Mains Road, Edinburgh EH9 3JF\\
  $^{9}$Institute for Computational Cosmology, Durham University, South Road, Durham DH1 3LE}
\begin{document}

\label{firstpage}

\maketitle

\begin{abstract} SCUBA-2 is an innovative 10000 pixel bolometer camera operating at submillimetre wavelengths on the James 
Clerk Maxwell Telescope (JCMT). The camera has the capability to carry out wide-field surveys to unprecedented depths, 
addressing key questions relating to the origins of galaxies, stars and planets. With two imaging arrays working 
simultaneously in the atmospheric windows at 450 and 850\,$\umu$m, the vast increase in pixel count means that SCUBA-2 maps 
the sky 100--150 times faster than the previous SCUBA instrument. In this paper we present an overview of the instrument, 
discuss the physical characteristics of the superconducting detector arrays, outline the observing modes and data acquisition, 
and present the early performance figures on the telescope. We also showcase the capabilities of the instrument via 
some early examples of the science SCUBA-2 has already undertaken. In February 2012, SCUBA-2 began a series of unique legacy 
surveys for the JCMT community. These surveys will take 2.5\,years and the results are already providing complementary data to 
the shorter wavelength, shallower, larger-area surveys from \emph{Herschel}. The SCUBA-2 surveys will also provide a wealth of 
information for further study with new facilities such as ALMA, and future telescopes such as CCAT and \emph{SPICA}. 
\end{abstract}


\section{Introduction}

The submillimetre waveband, which encompasses the spectral range from 0.3 to 1\,mm, contains a wealth of information about the 
cold Universe. Observations of gas and dust probe the earliest stages in the formation of galaxies, stars and planets. For 
example, the blackbody emission of a 10\,K source (or a 40\,K source at redshift $\sim$3) will peak around 300\,$\umu$m. The 
continuum emission from dust is usually optically thin, so observations can probe to the heart of the most crucial processes, 
with the consequence that, for example, embryonic star-forming core masses and the surrounding structure of their molecular 
clouds are determined in a less model-dependent way than in the optical and infrared 
\citep[e.g.][]{DiFrancesco2007,WardThompson2007}. On larger scales, much of the UV/optical light emitted from stars inside 
young galaxies is trapped within enshrouding dust clouds and re-emitted in the submillimetre. Only by observing at these 
longer wavelengths can the total energy budgets be determined. This is essential to derive an unbiased census of the star 
formation rate density with redshift and thus determine the ``formation epoch'' of galaxies 
\citep[e.g.][]{Blain1999,Murphy2011}.

\vskip 1mm

Undertaking submillimetre observations from ground-based observatories has always been fraught with difficulty, since 
atmospheric transparency is often poor and the high background power and sky emission variability limit the observing 
sensitivity. Nevertheless, 10--15\,m class single-dish telescopes, routinely operating with high efficiency for the past 25 
years, have led to enormous advances in our understanding of the formation of galaxies, stars and planets. For example, over 
the past two decades remarkable discoveries have taken place including the discovery of ultra-luminous high redshift galaxies 
responsible for the majority of the far-IR background \citep[e.g.][]{Smail1997,Hughes1998}, pin-pointing cold dense regions in 
molecular clouds where new stars are forming \citep[e.g.][]{Motte1998,Andre2010}, and imaging of vast clouds of cold dust 
around nearby stars believed to be analogues of the Kuiper Belt in our Solar System \citep[e.g.][]{Holland1998,Wyatt2008}.

\vskip 1mm

The submillimetre revolution began in earnest in the late 1990’s with the arrival of the first imaging cameras, SHARC 
\citep{Wang1996} on the Caltech Submillimeter Observatory telescope (CSO) and the Submillimetre Common-User Bolometer Array 
(SCUBA; \citealt{Holland1999}) on the JCMT. However, with two arrays containing only 91 and 37 bolometers, mapping even 
moderately-sized areas of sky (tens of arcminutes across) with SCUBA to any reasonable depth was painfully slow. Bolometer 
cameras with similar total pixel counts also followed on other ground-based telescopes, such as Bolocam \citep{Glenn1998} and 
SHARC-II \citep{Dowell2002} at the CSO, LABOCA \citep{Siringo2009} on the Atacama Pathfinder Experiment telescope, and the 
MAMBO cameras \citep{Kreysa1998} on the Institut de Radio Astronomie Millimetrique 30\,m telescope. With existing bolometer 
technology being non-scalable to more than a few hundred pixels, the next challenge was to develop a way to increase 
substantially the pixel count by up to a factor of 100. The solution came in the form of new detectors incorporating 
superconducting transition edge sensors (TES; \citealt{Irwin1995}), and the ability to adapt techniques such as 
high-precision silicon micro-machining to produce large-scale array structures \citep{Walton2005}. Furthermore, 
Superconducting Quantum Interference Device (SQUID) amplifiers could also be chained together to form a complementary, 
multiplexed readout system \citep{deKorte2003}. These technology advances meant that cameras of many thousands of pixels 
became conceivable for the first time, and thus formed the major motivation for the SCUBA-2 project.

\vskip 1mm

\begin{figure} 
\centering 
\includegraphics[width=90mm]{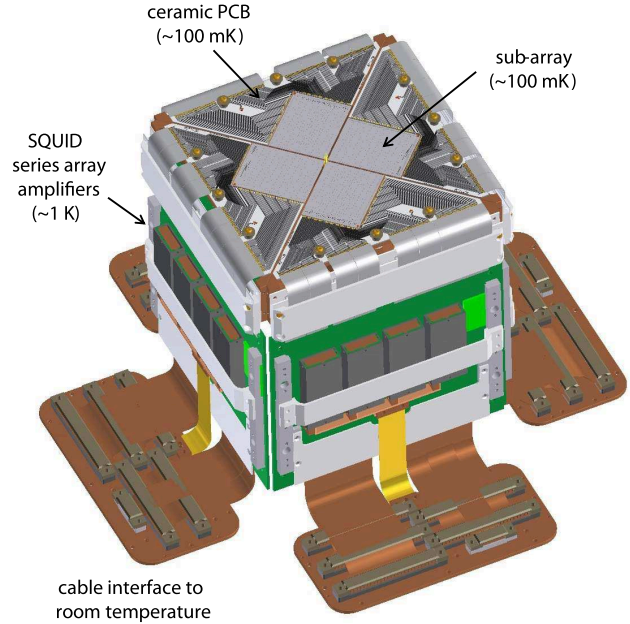} 

\caption{Layout of a SCUBA-2 focal plane unit showing the major components of the assembly. The sub-arrays are butted 
together, with an approximate 4 pixel gap, to form a focal plane with an approximate 45\,arcmin$^2$ field-of-view on the sky. 
Ceramic printed circuit boards (PCB) are wire-bonded to the arrays and these fan out the signal connections to ribbon cables 
that run to the 1\,K amplifiers and eventually via additional cables to room temperature.}

\label{fig:fplayout} 
\end{figure}

\begin{figure*} 
\centering 
\includegraphics[width=175mm]{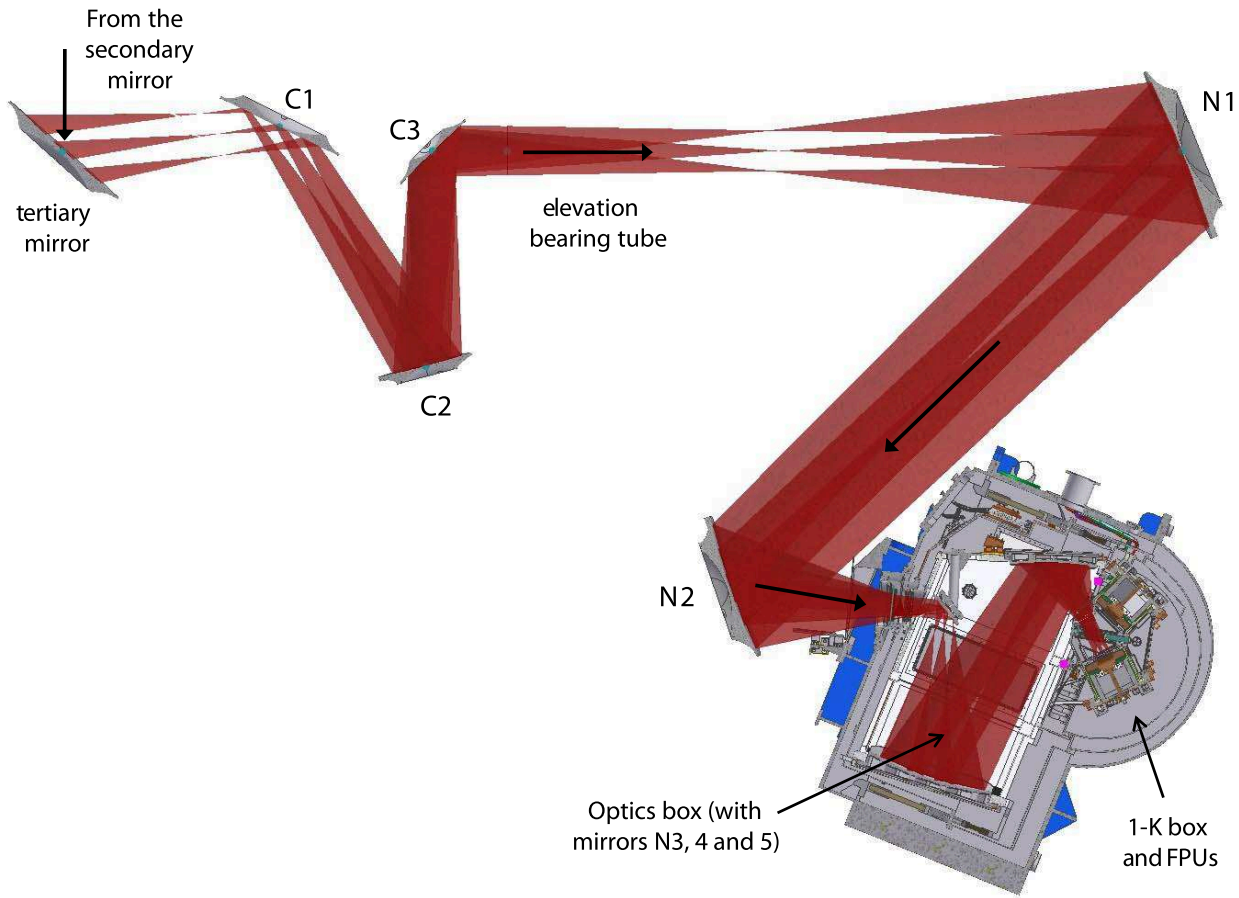} 

\caption{Optical layout for SCUBA-2 from the tertiary mirror to the detector arrays inside the cryostat. The beam envelope, 
shown in red, is a combined ray trace of the on-axis and two extremes of the field-of-view for this projection of the optics. The 
arrow shows the direction of light propagation. Mirror N3 is located just inside the cryostat window, whilst mirrors N4 and N5 
relay the optical beam into the array enclosure (``1-K box'') which houses the focal plane units (FPUs).}

\label{fig:optical_layout} 
\end{figure*}

SCUBA-2 is a dual-wavelength camera with 5120 pixels in each of two focal planes. A focal plane consists of 4 separate 
sub-arrays, each with 1280 bolometers, and butted together to give the full field (as shown in Fig. 1). Both focal planes have 
the same field-of-view on the sky and are used simultaneously by means of a dichroic beam-splitter. The instrument operates at 
the same primary wavelengths as SCUBA, namely 450\,$\umu$m for the short and 850\,$\umu$m for the long waveband. SCUBA-2 was 
delivered from the UK Astronomy Technology Centre (Edinburgh) to the Joint Astronomy Centre (Hilo, Hawaii) in 2008 April with 
one engineering-grade sub-array at each waveband. The first two science-grade sub-arrays (one for each focal plane) arrived at 
the JCMT in late 2009, and a period of ``shared risk observing'' was undertaken between February and April 2010. The 
remainder of the science-grade sub-arrays were delivered in summer 2010 and the first astronomical data with the full array 
complement were taken in early-2011. 

\vskip 1mm

In this paper Section 2 gives an overview of the instrument design including the optics and cryogenics. Section 3 describes 
in detail the design, manufacture and testing of the superconducting detector arrays. In Sections 4 and 5 we discuss how the 
instrument takes data and processes the information into astronomical images. Section 6 describes the rudiments of flux 
calibration whilst Section 7 presents the initial on-sky performance, including sensitivity and optical image quality. In 
Section 8 we give an overview of the algorithms used in reducing SCUBA-2 data to produce publication-quality images. Finally, 
Section 9 illustrates the scientific potential of SCUBA-2 with a selection of early results.

\section[]{Instrument design}

The opto-mechanical design of SCUBA-2 is driven by two principal requirements: (1) to maximise the available field-of-view; 
and, (2) to provide an ultra-low detector operating temperature in the 100\,mK regime. The re-imaging of a large field onto a 
relatively small detector array (Section 3.1.1), as well as infrastructure limitations at the telescope, results in a 
complex optical path necessitating some extremely large mirrors (up to 1.2\,m across). Furthermore, to minimise power loading 
on the detector arrays, the last 3 of the re-imaging mirrors are cooled to temperatures below 10\,K. Together with the complex 
cryogenic system, this leads to a large cryostat, the vacuum vessel of which is 2.3\,m high, 1.7\,m wide and 2.1\,m long, with 
a pumped volume of 5.3\,m$^3$ and a weight of 3400\,kg.

\subsection{Optical design}

Early designs clearly showed that it was not possible to accommodate SCUBA-2 in the JCMT ``receiver cabin'' close to the 
Cassegrain focus. The left-hand Nasmyth platform (as viewed from the rear of the telescope), previously home to SCUBA, was a 
more realistic location in which the unvignetted field-of-view of the JCMT is $\sim$11\,arcmin in diameter, restricted by the 
aperture of the elevation bearing. Given that the focal-plane has a square geometry (as dictated by the array manufacturing 
process), a maximum field of 8 $\times$ 8\,arcmin was possible. Hence the SCUBA-2 optics re-image the field at the Cassegrain 
focus to a size compatible with the focal plane footprint at the arrays. To maximise the sensitivity of the instrument and 
provide excellent image quality this has to be achieved with high efficiency and minimum field distortion. The optics are 
also designed to ensure that a high-quality pupil image of the secondary mirror is produced at a cold-stop within the 
cryostat thereby minimising ``stray light'' that could potentially degrade detector sensitivity. Subsequent changes to the 
array size (Section 3.1.1) restricts the final field-of-view to $\sim$45\,arcmin$^2$.

\vskip 1mm

The detailed optics design and manufacture of the re-imaging mirrors are described by \citet{Atad2006}. Referring to Fig. 2, 
the design consists of a tertiary mirror located in the receiver cabin just above the nominal Cassegrain focus. At the exit of 
the cabin a relay of three mirrors (labelled C1--C3) re-images the telescope focal plane at a point just beyond the elevation 
bearing on the Nasmyth platform, thereby converting the f/12 telescope beam to f/7. A second relay (N1 and N2) re-images the 
focal plane at f/2.7 just inside the instrument, thereby allowing for a small cryostat window diameter. The cold optics, 
consisting of a further 3 mirrors (N3, N4 and N5), forms an approximate 1:1 system that re-images the focal plane at f/2.7 
onto the detector arrays.

\vskip 1mm

The mirrors were manufactured by TNO Science and Industry\footnote{TNO Science and Industry, Opto-Mechanical Instrumentation, 
Precision Mechanics Department, Stieltjesweg 1, P. O. Box 155, Delft, The Netherlands.} to have complex free-form surfaces 
that provide sufficient degrees of freedom to optimise the optical design. This proved necessary to maintain a high Strehl 
ratio across the field as a function of telescope elevation, as well as minimum field distortion. Packaging the optics within 
the overall structure of the telescope results in a cryostat location just below the existing Nasymth platform and tilted at 
an angle of 22$^{\circ}$ to the vertical. This required a large amount of infra-structural changes at the telescope, as 
documented in \citet{Craig2010}. The overall optical path length is $\sim$20\,m from the tertiary mirror to the arrays. An 
alignment accuracy of $\pm$\,0.25\,mm, well within acceptable tolerances, is achieved in all axes using an optical datum 
positioned in the bearing tube \citep{Craig2010}.

\begin{figure} 
\centering 
\includegraphics[width=85mm]{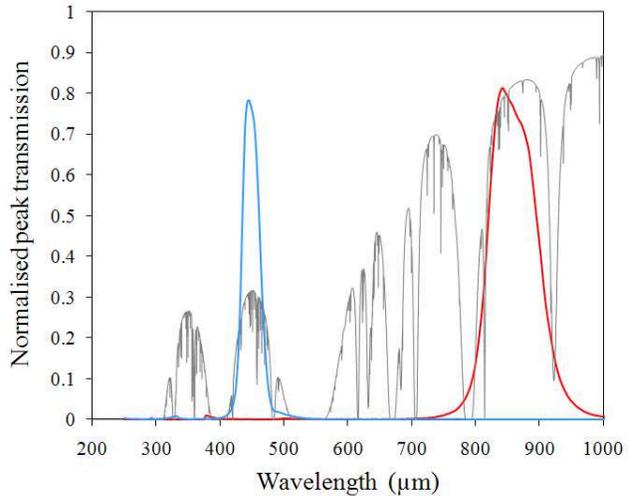} 

\caption{The measured SCUBA-2 bandpass filter profiles at 450 (blue curve) and 850\,$\umu$m (red), superimposed on the 
atmospheric transmission curve for Mauna Kea for 1\,mm of precipitable water vapour (PWV; grey curve). The atmospheric 
transmission data are provided courtesy of the Caltech Submillimeter Observatory.} 

\end{figure}

\subsection{Wavelength of operation}

Submillimetre observations from ground-based sites are restricted to wavebands within transmission windows in the atmosphere. 
For a good observing site such as Mauna Kea these windows extend from 300\,$\umu$m to 1\,mm and throughout the region 
atmospheric water vapour is the main absorber of radiation from astronomical sources. The selection of observing wavelength is 
made by a bandpass filter, which as shown in Fig. 3, is carefully tailored to match a particular transmission window. For 
SCUBA-2, these are multi-layer, metal-mesh interference filters \citep{Ade2006}, located just in front of the focal planes, 
and have excellent transmission (typically peaking around 80 per cent) and very low ($<$\,0.1 per cent) out-of-band power 
leakage. The half-power bandwidths of the bandpass filters are 32 and 85\,$\umu$m at 450 and 850\,$\umu$m, respectively, 
corresponding to $\lambda/\Delta\lambda$\,$\sim$\,14 and 10.

\vskip 1mm
 
A decision was made early in the design to conservatively filter the instrument. Hence there are a series of thermal and metal-mesh 
edge filters \citep{Tucker2006} to ensure that heat loads and stray light are kept to a minimum in such a large instrument. Fig. 4 
shows the position of all the filters within SCUBA-2, including the dichroic that reflects the shorter wavelengths and transmits the 
longer. The addition of extra low-pass edge filters at 4,K and 1,K is not a large penalty compared with potentially having to track 
down stray light sources that could mar image quality, or contending with additional heat loads that could degrade sensitivity. For 
example, at the entrance of the 4,K optics box it is necessary to keep the thermal power to a minimum to prevent heating of the optics 
and possible subsequent loading of the 1,K stage. To ensure good frequency selection low-pass edge filters are also used with the 
bandpass filters. The net transmission of the instrument, in both wavebands and including the cryostat window and the detector 
absorption efficiency, is $\sim$40 per cent.

\begin{figure} 
\centering 
\includegraphics[width=85mm]{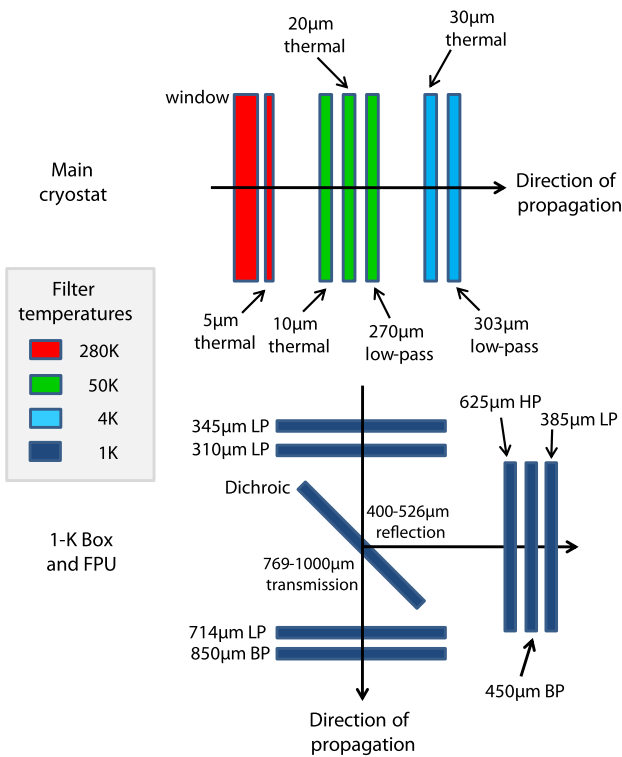} 

\caption{The arrangement and operating temperature (colour coded) of the bandpass (BP), thermal blocking and metal-mesh edge 
filters and dichroic in SCUBA-2. Top: the main cryostat; Bottom: the 1-K box and focal plane units. ``LP'' and ``HP'' 
represent low-pass and high-pass filter cut-off edges, respectively. Thermal edge filters reject power shortward of their 
wavelength cut-off.}

\label{fig:filter_stack} 
\end{figure}

\subsection{Main instrument}

The cryostat is made up of a series of sub-systems (see Fig. 2) and is designed with nested radiation shields and baffles to 
minimise stray light and magnetic fields. For example, the arrays themselves contain SQUID amplifiers (Section 3.2.2) that 
are sensitive magnetometers and so must be shielded from magnetic fields. Immediately inside the vacuum vessel is a high 
magnetic permeability shield, a multi-layer insulation blanket and a radiation shield operating at $\sim$50\,K. These 
provide radiation shielding for the main optics box, that houses the cold re-imaging mirrors at $\sim$4\,K. The radiation 
shield and optics box are cooled by a pair of pulse-tube coolers (Section 2.5). The main optics box provides the support 
for the three cold mirrors and the 1\,K enclosure (``1-K box''). Mounted within the 1-K box are the two focal plane units 
(FPUs) that contain the cold electronics and the detector arrays. The still and the mixing chamber of a dilution refrigerator 
(DR) cool the 1-K box and arrays respectively (Section 2.5). The 1-K box and the outer casing of each FPU are also 
wrapped in superconducting and high magnetic permeability material \citep{HollisterMagSh,Craig2010}.

\subsection{1-K box and focal plane units}

The removable 1-K box creates the required environment for the detector arrays \citep{Woodcraft2009}. In addition to radiation 
shielding, it provides a cold-stop aperture at the entrance to help minimise stray light. Furthermore, it gives mechanical 
support for magnetic shielding, a cold shutter (used to take dark frames), filters, and the dichroic that splits the incoming 
beam onto the two focal planes. The 1-K box consists of an outer shell with aluminium alloy panels that hold the high 
permeability material for magnetic shielding \citep{HollisterMagSh}. In addition, the box accurately and reproducibly supports 
and positions the FPUs with respect to the cold-stop. Fig. 5 (left) shows a 3-D CAD drawing of the box, highlighting the main 
components. There are two separate focal plane units, each containing four sub-arrays (Section 3.1.1). Key elements of the 
FPU design include the thermal link to the DR, optical filtering and further magnetic shielding. The 1-K box is a separate 
sub-system and interfaces to the main cryostat assembly via a support frame. Fig. 5 (right) shows a photograph of the fully 
assembled 1-K box during installation into the instrument.

\begin{figure*} 
\centering 
\includegraphics[width=175mm]{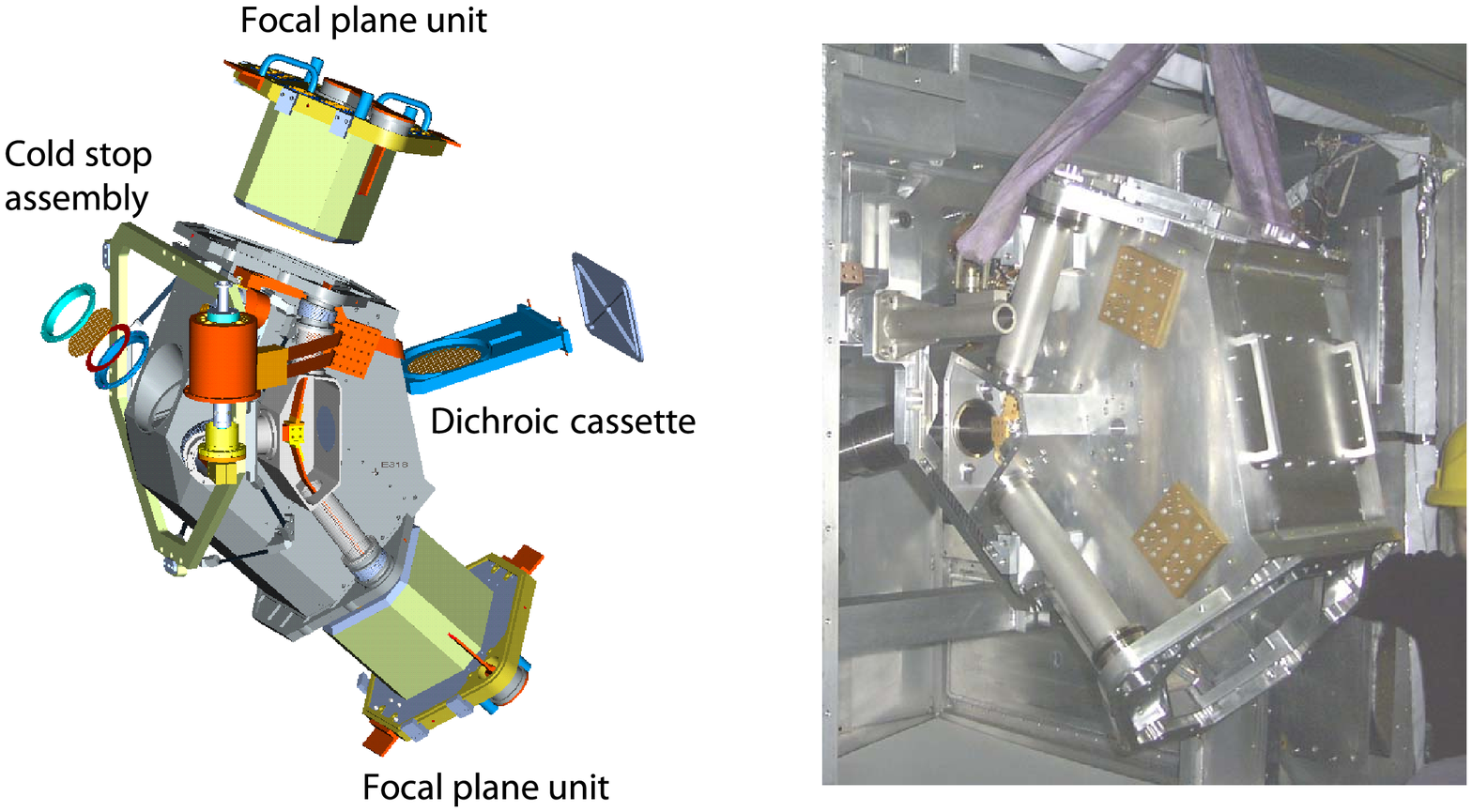} 

\caption{Left: 3-D CAD drawing of the 1-K box showing the main components of the focal-plane units, cold-stop assembly and 
dichroic cassette. The shutter, not shown in this drawing, is located in front of the cold-stop on the outside of the 1-K box; 
Right: Photograph of the fully assembled 1-K box during installation into the main SCUBA-2 cryostat.}

\label{fig:1kbox} 
\end{figure*}

\subsection{Thermal design and cryogenics}

The overall thermal design of SCUBA-2 is described by \citet{Gostick2004}. In summary, two Cryomech\footnote{Cryomech, 113 
Falso Drive, Syracuse, NY13211, USA.} PT410 pulse-tube coolers keep the radiation shields and the $\sim$300\,kg of cold optics 
at 50 and 4\,K, respectively. However, since their cooling power is insufficient for the initial cool-down phase on a 
reasonable timescale, pre-cool tanks are attached to the 50 and 4\,K shields. After pre-cooling with liquid nitrogen (LN$_2$) 
the instrument is kept cold without the need for any liquid cryogens in the main instrument. A modified Leiden 
Cryogenics\footnote{Leiden Cryogenics BV, Galgewater 21, 2311 VZ Leiden, The Netherlands.} dilution refrigerator was 
commissioned to run with a pulse-tube cooler (PT410) and Joule-Thompson heat exchanger, eliminating the need for a 
conventional 1\,K pot and liquid cryogens. The still of the DR cools the 1-K box, whilst the mixing chamber cools the 
$\sim$30\,kg of focal planes to around 100\,mK. The DR is a key element of the system and has to cope with a substantial 
thermal load from the arrays themselves, heat leaks down the mechanical array supports and wiring, as well as radiation 
loading from the warmer parts of the instrument, telescope and sky \citep{HollisterDR}. The thermal design is complex, with 
the need to transfer cooling power at temperatures of 1\,K and 100\,mK over a distance of 1.5\,m to various locations in the 
FPUs, and to support the arrays rigidly whilst keeping sufficient thermal isolation to the 100\,mK stage. A large number of 
thermal links are therefore required, with the added need for several bolted interfaces to allow the FPUs to be removed from 
the instrument. Nevertheless, with the benefit of extensive thermal modelling, the instrument reached the required cryogenic 
performance on the first cool-down. Under a total thermal load of 70\,$\umu$W the mixing chamber of the DR achieves a base 
temperature of 70\,mK in regular operation \citep{Bintley2012a}.

\vskip 1mm

Early operation on the telescope revealed two main problems. The first was that the DR was prone to blocking after only 4--5 
weeks of continuous operation. This was due to a gradual build-up of contamination not removed by the LN$_2$ cold traps, which 
over time causes a blockage, most likely in the flow impedance of the cold insert. An additional external 4\,K cold-trap 
(cooled by liquid helium) significantly extended the run-time, allowing the instrument to remain cold for more than 6 months 
continuously. The second issue was that a very distinct oscillation (period $\sim\,$25\,s) was seen in the bolometer output 
signals. This was traced to a temperature oscillation originating in the still of the DR. The oscillation is a result of both 
tilting the DR to 22$^{\circ}$ and the strong interaction between the still and the circulation of $^3$He gas (this being a 
consequence of using the still to condense the gas as part of the new DR design; \citealt{Bintley2012a}). A new temperature 
control system on both the support structure underneath the arrays and 1-K box minimised the amplitude of the oscillation. The 
temperature fluctuations have been reduced by at least a factor of 10, to $\pm$\,20\,$\umu$K at the array supports. Under 
temperature control the mixing chamber achieves a base temperature of 78\,mK.

\section{Detector arrays}

\subsection{Requirements}

\subsubsection{Pixel count and array geometry}

To fully Nyquist sample the sky instantaneously, the detector spacing must be $0.5\,f\lambda$, where $f$ is the final focal 
ratio of the optics. With f/2.7 this corresponds to a spacing (and approximate detector diameter) of 0.61 and 1.14\,mm at 450 
and 850\,$\umu$m, respectively. To cover the maximum available field-of-view requires approximately 25,600 and 6,400 
bolometers for the two wavebands. Early design work showed that there was an approximate 1.1\,mm minimum constraint on the 
size scale of the multiplexer (MUX) unit cell, rendering a fully-sampled 450\,$\umu$m focal plane impractical. The detector 
size and spacing was therefore relaxed to $\sim$$f\lambda$ at 450\,$\umu$m producing an array that under-samples the sky by a 
factor of 4, leading to a subsequent reduction in mapping speed (Section 7.3). However, this decision greatly simplified the 
fabrication process, since the multiplexer wafers became identical at the two wavelengths (Section 3.2.2). Furthermore, 
fabrication limitations meant that the maximum size of an individual detector or MUX wafer was 50\,mm$^2$, and hence the 
focal planes are populated with four separate quadrants, or sub-arrays. Finally, the need for space on the MUX wafer for 
wire-bond pads, extra bump-bonds, and the second stage SQUID configuration, means that the size of a sub-array is further 
restricted to 32 columns by 40 active rows (Section 3.2.2).

\begin{table*}
 \centering
  \caption{Summary of the predicted power levels under the best and worst atmospheric conditions, and the per-bolometer 
requirements in terms of power handling, NEP, transition temperature, thermal conductance and time constant. The minimum 
background power is estimated in the best observing conditions for which zenith sky transmissions of 40 and 85 per cent at 450 and 
850\,$\umu$m have been adopted. For the maximum power levels transmission values of 3 and 40 per cent have been used.}
  \begin{tabular}{@{}llcc@{}}
  \hline

   Parameter                               &  Units            & 450\,$\umu$m                  & 850\,$\umu$m                   \\
                                                               &                               &                                \\
 \hline
   Minimum background power                & (pW)              & 70                            & 7                              \\
 
   Maximum background power                & (pW)              & 120                           & 16                             \\

   Total power handling/saturation power   & (pW)              & 230 ($\pm$\,10 per cent)      & 50 ($\pm$\,10 per cent)        \\

   Minimum background NEP                  & (W\,s$^{1/2}$)    & 2.7\,$\times$\,10$^{-16}$     & 5.6\,$\times$\,10$^{-17}$      \\

   Detector (phonon) noise limited NEP     & (W\,s$^{1/2}$)    & $<$1.35\,$\times$\,10$^{-16}$ & $<$2.8\,$\times$\,10$^{-17}$   \\

   Measured dark NEP                       & (W\,s$^{1/2}$)    & $<$1.9\,$\times$\,10$^{-16}$  & $<$4.0\,$\times$\,10$^{-17}$   \\

   Transition temperature (T$_{\rmn{c}}$)  & (mK)              & 190 ($\pm$\,5 per cent)       & 130 ($\pm$\,5 per cent)        \\   
  
   Thermal conductance ($G$)               & (nW\,K$^{-1}$)    & 4.2 ($\pm$\,5 per cent)       & 1.3 ($\pm$\,5 per cent)        \\

   Time constant                           & (ms)              & $<$\,1.5                      & $<$\,2.8                       \\

\hline
\end{tabular}
\end{table*}

\vskip 1mm

\subsubsection{NEP and power handling requirements}

The key performance requirements for the detectors are the bolometer noise (or noise equivalent power, NEP; Section 
3.5.2), the power handling capability (saturation power), and the speed of response (time constant). The fundamental 
requirement is that the overall SCUBA-2 sensitivity be limited by the background photon noise due to sky, telescope and 
instrument, under all observing conditions. A detailed model, based on the heritage of SCUBA, was constructed to allow the 
background power levels and performance figures to be established. As shown in Table 1, the 450\,$\umu$m bolometers have to 
cope with larger sky power levels than their 850\,$\umu$m counterparts, and under the driest observing conditions the 
background power is approximately 10 times less at 850\,$\umu$m than at 450\,$\umu$m. The specification of the total power handling 
capability therefore takes this into account, and also has to include additional margin for electrical TES bias and the 
calibration heater power (Section 3.4.1). In terms of the NEP, the 850\,$\umu$m waveband sets the most stringent 
requirement as the sky background power is considerably lower than at 450\,$\umu$m. The minimum background-limited NEPs 
(${\rmn{NEP}}_{\rmn{bkg}}$) are 2.7 $\times$ 10$^{-16}$ and 5.6 $\times$ 10$^{-17}$\,W\,s$^{1/2}$ at 450 and 850\,$\umu$m, 
respectively. Hence, the intrinsic NEP of a bolometer must be less than these values to be background limited. For an ideal 
TES bolometer, measured in the absence of background power, phonon noise dominates the NEP at low frequencies (Section 
3.2.1). Hence, the specification adopted for SCUBA-2 is that the phonon noise limited NEP for an individual bolometer is 
$<$\,0.5 $\times$ ${\rmn{NEP}}_{\rmn{bkg}}$. Given that the NEP will be degraded by additional noise in the readout circuit 
the formal specification is that the measured \emph{dark} NEP (i.e. measured in the absence of background power) is $<$\,0.7 
$\times$ ${\rmn{NEP}}_{\rmn{bkg}}$. These values are summarised in Table 1.

\vskip 1mm

\subsubsection{Frequency response}

SCUBA-2 is designed to conduct large-area surveys by scanning the telescope in a rapid, overlapping pattern (Section 5.1). If 
the detector response is too slow, some of the higher spatial frequencies in the science signal will be attenuated. The 
telescope acts as a spatial filter, since the measured response is the convolution of the response of the astronomical signal 
and the telescope beam. The maximum frequency present in the system response is given by $v_{\rmn{tel}}$/$(p f \lambda$) where 
$v_{\rmn{tel}}$ is the telescope scanning speed and $p$ the plate scale (5 arcsec\,mm$^{-1}$). Since the telescope can scan at 
speeds up to 600 arcsec\,sec$^{-1}$ with high positional accuracy, resulting data will therefore have maximum frequencies 
present of 100 and 50\,Hz at 450 and 850\,$\umu$m, respectively. Thus the detector time constants must be $<$\,1.5 and 
$<$\,2.8\,ms to avoid significant attentuation of the signal during fast scanning. More details of the derivation of the 
detector and array requirements are given in \citet{Hollister2009}.

\subsection{Design and fabrication}

The SCUBA-2 sub-arrays are based on transition edge sensors and time-division SQUID-based multiplexers developed at the 
National Institute of Standards and Technology in Boulder, Colorado \citep{Irwin2005}. The superconducting elements themselves 
are formed from a molybdenum/copper (Mo/Cu) bilayer of material, the relative thickness of each layer determining the 
superconducting transition temperature. The geometry of each sub-array is 32 columns by 40 active rows of bolometers. As shown 
in the schematic diagram of a single bolometer in Fig. 6, each sub-array consists of two separate wafers, fabricated 
separately and then hybridised together.

\subsubsection{Detector wafer}

The ''detector wafer'' upper surface is implanted with phosphorus ions to provide an absorbing layer to incoming 
electromagnetic radiation, whilst the Mo/Cu bilayer on the lower surface forms a highly reflective backshort. For efficient 
radiation absorption the thickness of the wafer is made equal to an odd number of quarter wavelengths, 3$\lambda$/4 at 
450\,$\umu$m and $\lambda$/4 at 850\,$\umu$m \citep{Audley2004}. An underside silicon nitride membrane mechanically supports 
each TES bolometer on the wafer and provides a weak thermal link to the cold bath. For ease of manufacture the sub-arrays for 
450 and 850\,$\umu$m are identical, except for the aforementioned thickness of the detector wafer. The final part of the 
detector wafer design is a heater circuit arranged in a thin-line geometry around the edge of each bolometer (Section 3.4.2). 
In terms of fabrication, and to ease handling, a second silicon wafer is fusion bonded to the upper surface of the detector 
wafer (after the implantation stage). This is eventually removed after hybridisation with the MUX wafer, just prior to the 
post-processing step that thermally isolates each bolometer via a deep-etched 10\,$\umu$m wide trench to the silicon nitride 
layer (Section 3.2.3).

\vskip 1mm

The detector operating temperature and thermal conductance of the link to the cold bath govern the theoretically achievable NEP, 
according to ${\rmn{NEP}}_{\rmn{phonon}} = \sqrt{4 \gamma k_{\rmn{B}}T^2 G}$, where $T$ is operating temperature (approximated by 
the superconducting transition temperature, $T_{\rmn{c}}$), $G$ the thermal conductance and $\gamma$ is a factor that accounts for 
the temperature gradient across the silicon nitride membrane (assumed to be 0.7 in this case; \citealt{Mather1982}). Target values 
of $T_{\rmn{c}}$ and $G$ are chosen to provide background limited performance, and also to contend with varying degrees of power as 
the sky emission changes. As discussed in Section 3.1.2 the 850\,$\umu$m waveband sets the most stringent requirement in terms of 
NEP. In addition, to minimise sensitivity to temperature fluctuations $T_{\rmn{c}}$ should be roughly twice the expected base 
temperature. At 850\,$\umu$m the value of $T_{\rmn{c}}$ is therefore set to 130\,mK. At 450\,$\umu$m, where the sky power is 
higher, $T_{\rmn{c}}$ can also be made higher, in line with the relaxed NEP requirement. Strictly, only a value of 380\,mK is 
needed to ensure the 450\,$\umu$m waveband is background limited. However, given the desire to keep the array fabrication common to 
both wavebands, and that a 100\,mK operating temperature regime is needed in any case for 850\,$\umu$m, a $T_{\rmn{c}}$ of 190\,mK 
was adopted at 450\,$\umu$m, thereby giving even more margin on the required NEP. The value of $G$ is given by $G \sim (nP)/T$, 
where $P$ is the saturation power and $n$ is a power-law constant, which is typically 3.5 for silicon nitride membranes. Hence, the 
required values of $G$ are 4.2 and 1.3\,nW\,K$^{-1}$ at 450 and 850\,$\umu$m, respectively. The target $T_{\rmn{c}}$ and G values 
are given in Table 1.

\begin{figure} 
\centering 
\includegraphics[width=85mm]{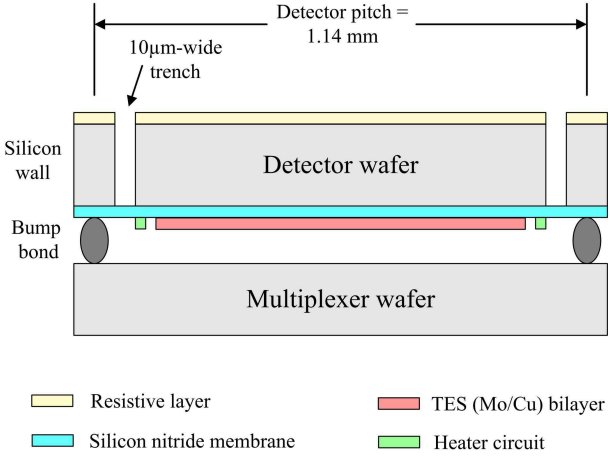} 

\caption{A schematic representation of a single SCUBA-2 bolometer showing the principal components. These include the 
absorbing resistive layer on the top of the detector wafer and the deep trenches that thermally isolate each bolometer from 
their neighbours. The TES bilayer sits between the silicon nitride membrane and the bottom of the detector wafer. The heater 
circuit runs around the edge of each bolometer and is used for calibration. The multiplexer wafer containing the SQUID 
amplifer circuitry is indium bump-bonded to the detector wafer. Note: Diagram is not to scale. }

\label{fig:pixel} 
\end{figure}

\begin{figure} 
\centering 
\includegraphics[width=85mm]{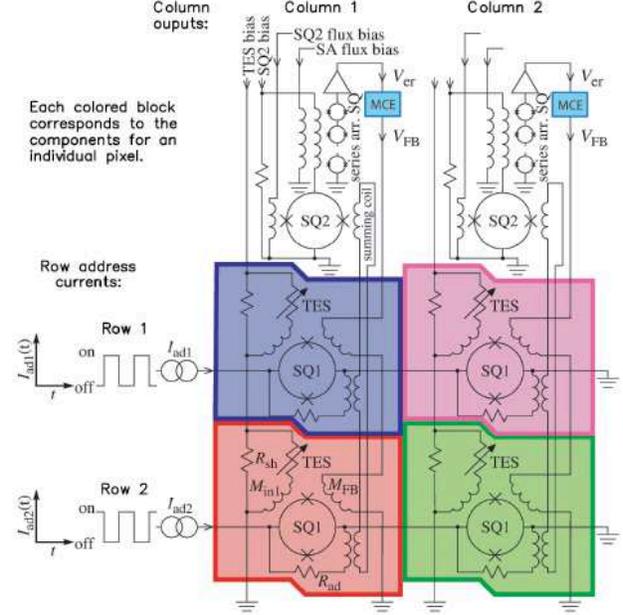} 

\caption{A two-column, two-row schematic representation of the SQUID time-division multiplexed readout scheme for SCUBA-2 
\citep{Doriese2007}. Each TES is inductively coupled to its own first-stage SQUID (SQ1). A summing coil carries the signals 
from the SQ1s in a column to a common second-stage SQUID (SQ2). Rows of SQ1s are sequentially switched on using an address 
current ($I_{ad}$), so the signal from one TES at a time (per column) is passed to the SQ2. Finally, the output of each SQ2 is 
passed to a 100 SQUID series array amplifer (SSA) and then to the room temperature electronics (MCE). }

\label{fig:pixel} 
\end{figure}

\vskip 1mm

\subsubsection{Multiplexer wafer}

The bottom or ``multiplexer'' wafer contains the input coupling coils and SQUID amplifiers of the readout circuit (as shown in Fig. 
7). Current flowing through the TES generates a magnetic field at the first-stage SQUID (SQ1) through an input transformer on the 
wafer (the details of which are not shown in Fig. 7 for clarity). Each column of SQ1s is then coupled by a summing coil to the 
second stage SQUID (SQ2). The signals are amplified by a SQUID Series Array (SSA) that has 100 SQUIDS in series per channel and is 
located on the 1\,K PCB of the cold electronics module (see Section 3.3). The MUX wafer is designed with 41 rows with the first row 
being a ``dark row'', without any corresponding TES element but containing a SQ1. The output of the dark row has been used to 
investigate common-mode noise on a per-column basis, but is not currently implemented by default in the data reduction software. It 
is the SQUID MUX that makes large-scale TES arrays, such as SCUBA-2, practical by vastly reducing the wire count between the 
detectors and the room temperature electronics \citep{deKorte2003}. The SCUBA-2 MUX design reduces the wire count from 82000 
to 2700 -- a MUX factor of approximately 30. The MUX wafers are independently tested prior to hybridisation with the detector 
wafer, using a dedicated facility at the University of Waterloo that measures yield and critical currents ($I_{\rmn{c}}$) of the 
MUX wafers. Testing at this stage allows fabrication faults to be identified and corrected.

\vskip 1mm

\subsubsection{Hybridisation and post-processing}

The detector and multiplexer wafers are hybridised together using a low-temperature indium bump-bonding process developed at 
Raytheon Vision Systems\footnote{Raytheon Vision Systems, 74 Coromar Drive, Goleta, California 93117, USA.}. The bump-bonds 
provide both thermal and electrical contact between the two wafers. There are 74 bumps surrounding each detector element 
(including 4 bumps that make the electrical connection between wafers for the bias and heater) and a further 100,000 bumps per 
sub-array around the perimeter of the wafers to give extra mechanical support. The first step of post-processing is to etch 
away the ``handle wafer'' to the level of the implanted absorbing layer. This is followed by thermally isolating each 
individual bolometer by deep etching a trench in the main detector wafer to the silicon nitride membrane (see Fig. 6; 
\citealt{Walton2005}). The trenches are 10\,$\umu$m wide and either 60 or 100\,$\umu$m deep depending on the thickness of the 
detector wafer (Section 3.2.1). Maintaining this width at the bottom of the trench across the entire sub-array is critical as 
this (largely) controls the value of $G$. At this stage a final electrical continuity check allows any remaining fabrication 
issues to be repaired (such as electrical shorts that may have been introduced in the hybridisation process). The final step 
in the array processing is to laser dice the circular wafer assembly into the final rectangular sub-array geometry.

\begin{figure} 
\centering 
\includegraphics[width=85mm]{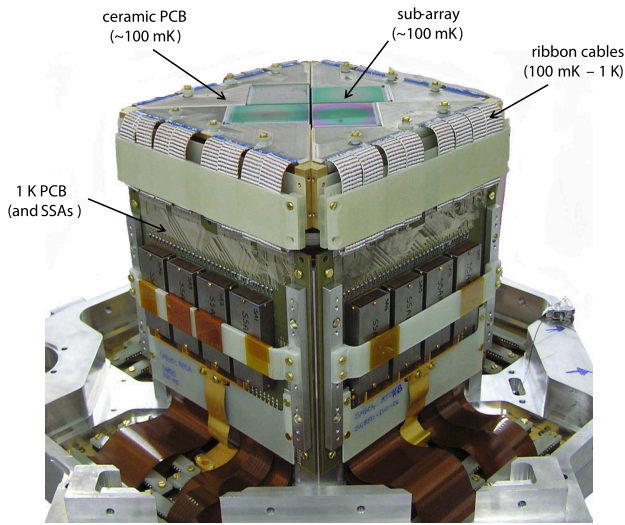} 

\caption{Photograph of four sub-array modules folded into position in a focal plane unit. The principal components are 
highlighted.}

\label{fig:module} 
\end{figure}

\subsection{Array integration}

Completed sub-arrays are packaged as stand-alone modules that are mounted in the focal plane of the instrument. The sub-array 
is first epoxy-bonded onto an array support holder. As thermal conduction laterally through the sub-array is poor, the holder 
needs to make thermal contact to the entire back surface of the sub-array to provide sufficient cooling. The holder must also 
be made from a metal for effective thermal conduction, but this results in a large mismatch in thermal contraction between the 
holder and the (largely silicon) sub-array. The array holder is therefore designed in the form of a beryllium-copper block in 
which individual spark-eroded tines make contact with the underside of the MUX wafer through an epoxy bond, with the pitch of 
the tines being identical to that of the MUX unit cells. By allowing for differential thermal contraction during cooling 
damage to the sub-array is avoided. Once attached to this ``hairbrush'' array holder the sub-array is integrated into the 
``sub-array module'', making electrical connection to a ceramic PCB (see Fig. 1) through aluminium wire bonds. Phosphor 
bronze-clad niobium titanium (NbTi) wires, woven into Nomex\textregistered\, cables (manufactured by Tekdata 
Interconnections\footnote{Tekdata Interconnections Ltd., Innovation House, The Glades, Festival Way, Etruria, Stoke-on-Trent, 
Staffordshire ST1 5SQ, UK.}), carry the signals from the ceramic PCB to a 1\,K PCB, that houses the magnetically shielded 
SSAs. Further woven ribbon cables (monel-coated NbTi) take the signals to the warm electronics on the outside of the cryostat. 
The design of both sets of cable is critical to minimise any heat leaks from either 1\,K or higher temperatures. Fig. 8 shows 
a photograph of 4 sub-array modules folded into position in a FPU.

\subsection{Array operation}

\subsubsection{Sub-array set up and bias optimisation}

Before operation can begin the arrays must be set up in their optimum configuration. This process has three main steps, the 
first two of which are performed quite rarely as the parameters are fixed and unlikely to vary with time (at least on a per 
cool-down basis). The first stage is called ``full array setup'' and refers to the process of determining the optimal SQUID 
bias for each level of SQUIDs. It sets the SSA bias to $I_{\rmn{c}}$(max) for maximum modulation, second stage SQUID bias to 
1.5–-2 $I_{\rmn{c}}$(max) for optimal bandwidth, and the first-stage SQUID bias to the mode value of the 32 bias settings that 
gave maximum modulation of the SQ1 for each row. The second stage or ``detector setup'' refers to the process of selecting the 
optimal TES bolometer and heater biases for each array. This involves sweeping out the available parameter space and selecting 
operating values such that the NEP across a sub-array is minimised. The final step, and the one that is performed regularly, 
is ``fast array setup'' and refers to the process of determining the flux offsets for each level of SQUIDs with the SQUID, TES 
and heater biases set to their nominal operating values. The array setup process is more fully described in \citet{Gao2008}.

\subsubsection{Heater tracking}

One of the innovative features of SCUBA-2 is the inclusion of a resistive heater arranged around the edge of every bolometer.  
The heaters play a fundamental role in the operation during observing in that they are used to compensate for changes in 
optical power as the sky background changes, enabling the TES bias point to be constant for a wide range of sky powers. 
Furthermore, each bolometer is individually calibrated by measuring its responsivity using a small ramp of the heater current 
(Section 5.2). The optical power from the sky is directly measured using a process called ``heater tracking''. This 
involves running a servo loop on the heater to keep the bolometer output constant while opening and closing the cold shutter 
to the sky. Periodic heater tracking transfers the slow changes in sky power to the heater setting, thereby maintaining the 
optimal power balance in a bolometer as determined during the array setup. The absolute level of power depends on the heater 
resistor values. In practice the average heater current from approximately 100 of the most stable bolometers on a sub-array 
are monitored. Although the resistors are nominally 3\,$\Omega$, all the power from the resistors is not necessarily 
coupled to the TES film. Thus each sub-array has a ``heater coupling efficiency factor'' (Section 3.5.1) to ensure that 
the responsivity and hence the NEP (Section 3.5.2) is well-calibrated.

\begin{table*}
 \centering
  \caption{The measured thermal and electrical properties of the SCUBA-2 science-grade sub-arrays, including a comparison of 
the expected phonon noise limit with the measured NEP in the dark. The naming convention for the sub-arrays is s4a, s4b, 
s4c and s4d for the 450\,$\umu$m focal plane, and s8a, s8b, s8c and s8d at 850\,$\umu$m.}
  \begin{tabular}{@{}lcccccc@{}}
  \hline

   Sub-array & Measured $G$   & Measured $T_{\rmn{c}}$ & Phonon-limited     & Saturation  & Measured                 & Yield       \\
   name      &                &                 & NEP                       & power       & dark NEP                 &             \\
             & (nW\,K$^{-1}$) & (mK)            & (W\,s$^{1/2}$)          & (pW)        & (W\,s$^{1/2}$)         & (per cent)  \\
 \hline
   s4a       & 4.9            & 212             & 9.2  $\times$ 10$^{-17}$  & 328         & 3.2 $\times$ 10$^{-16}$  &   84        \\
   s4b       & 6.1            & 206             & 1.0  $\times$ 10$^{-16}$  & 356         & 2.7 $\times$ 10$^{-16}$  &   72        \\
   s4c       & 8.5            & 203             & 1.2  $\times$ 10$^{-16}$  & 541         & 4.6 $\times$ 10$^{-16}$  &   65        \\
   s4d       & 6.1            & 198             & 9.6  $\times$ 10$^{-17}$  & 372         & 2.7 $\times$ 10$^{-16}$  &   65        \\
   s8a       & 4.3            & 145             & 5.9  $\times$ 10$^{-17}$  & 162         & 1.1 $\times$ 10$^{-16}$  &   80        \\
   s8b       & 2.8            & 130             & 4.3  $\times$ 10$^{-17}$  & 87          & 1.5 $\times$ 10$^{-16}$  &   60        \\
   s8c       & 3.7            & 154             & 5.3  $\times$ 10$^{-17}$  & 162         & 1.1 $\times$ 10$^{-16}$  &   61        \\
   s8d       & 5.7            & 147             & 5.8  $\times$ 10$^{-17}$  & 238         & 1.6 $\times$ 10$^{-16}$  &   65        \\
\hline
\end{tabular}
\end{table*}

\subsubsection{Sub-array operation}

The SCUBA-2 TES bolometers are operated in an approximate voltage-biased mode using a small 5\,m$\Omega$ shunt resistor 
($R_{\rmn{sh}}$) located on the MUX wafer (as shown in Fig. 7). The advantage of voltage-biasing the TES is that negative 
electro-thermal feedback (ETF) stabilises the bolometer against thermal runaway. An increase in background power warms the device 
and causes an increase in resistance, which in turn causes the bolometer current to decrease, thereby cooling the TES. Strong ETF 
essentially keeps the temperature of the TES constant, while providing a simple and direct relation between any applied power 
(optical or heater) and the current flowing through the device. Negative feedback also makes the bolometer self-biasing in terms of 
temperature in the transition. Variations in the incident power are automatically compensated for by changes in the bias current 
power on timescales shorter than the time-constant of the bolometer and via the heater for longer-term drifts (Section 3.4.2). As 
with all such devices, with too much applied power (optical, thermal or electrical bias) the TES becomes normal and ceases to work 
as a bolometer, and with too little applied power the TES becomes superconducting, with the same effect.

\vskip 1mm

The current flowing through each TES element is measured by its own first stage SQUID (SQ1). The output of a SQUID is periodic 
with magnetic flux from the input coil, the periodicity being given by a flux quantum \citep{deKorte2003}. Since there is no 
unique output for a given detector current the SQ1 is used as a null detector. Current is applied by the room temperature 
electronics (Section 4.1) to the SQ1 feedback coil to null the field from the TES current in the input coil. By applying a 
flux locked loop, the applied feedback current is proportional to the current flowing through the TES. The dynamic range of 
the detector feedback circuit is limited by the available first stage SQUID feedback current and the mutual inductance of the 
SQ1 input coil. These parameters are carefully chosen to meet the stringent noise requirements of the instrument.

\subsection{Sub-array performance}

The first two science-grade sub-arrays were tested individually in a dedicated cryostat at Cardiff University 
\citep{Bintley2010}. All of the sub-arrays were then either re-tested or tested for the first time in the SCUBA-2 instrument 
at the telescope. This aimed to characterise the sub-array performance initially under dark conditions (i.e. with the shutter 
closed; as presented in this section) and then on the sky under observing conditions (Section 7.2). The power leakage 
around the shutter when closed is small ($<$\,0.5\,pW) compared with, for example, a minimum sky power of $\sim$7\,pW at 
850\,$\umu$m.

\subsubsection{Thermal and electrical characteristics}

As discussed in Section 3.2.1 the operating (and transition) temperature and the thermal conductance to the cold bath dictate 
the achievable detector NEP and control the total power handling capability. The measurement of $T_{\rmn{c}}$ and $G$ starts 
with the bolometers in the normal state. The heater current is gradually reduced until the TES passes through its transition, 
with a small amount of bias power helping to identify the start of the transition. This process is then repeated at different 
temperatures. The measurement technique requires an accurate calibration of the heater resistance. As discussed in Section 
3.4.2, the ``effective'' heater resistance will be lower than the design value because of the imperfect coupling between the 
heater and the TES element, and inevitably some heat will flow into the walls between bolometers. The effective resistance is 
determined from a series of I--V curves at different heater settings. The response of each sub-array is then normalised by a 
``heater coupling efficiency'' factor based on optical measurements performed with ambient and LN$_2$ temperature loads at the 
window of the cryostat and by observation of standard calibration sources (Section 6.2). This ensures that each sub-array 
reports equal power when observing the same source. Table 2 gives the mean $T_{\rmn{c}}$ and $G$ across each of the 8 
sub-arrays. The detector time constants are measured by applying a square wave function to the heater and measuring the 
bolometer response using a fast readout mode available with the room temperature electronics. The measured time-constants are 
typically 1\,ms.

\vskip 1mm

Although test ``witness'' samples were taken during the $T_{\rmn{c}}$ deposition processes, the measured values of 
$T_{\rmn{c}}$ are, in most cases 8--10 per cent higher than the specification of 190\,($\pm$\,5) and 130\,($\pm$\,5)\,mK for 
the 450 and 850\,$\umu$m bolometers respectively. The higher-than-expected $T_{\rmn{c}}$ is not well understood but one 
possibility is that the wafers suffered from annealing in the processing after the bilayer deposition stage. The variation in 
$T_{\rmn{c}}$ on an individual sub-array is mainly radial, with values being lower in the (offset) centre position and 
typically increasing by 10 per cent towards the edges \citep{Bintley2012b}. This is a consequence of the sputtering process in 
which the detector wafer spins as the copper and molybdenum are deposited.

\vskip 1mm

The values of $G$ are much higher than the design, typically by factors of 2--3. $G$ tends to be more uniform across the array 
although there is a slight radial dependence similar to $T_{\rmn{c}}$, being smaller in magnitude towards the centre. The 
reason why $G$ is so much higher than the requirement is not well understood. Sub-array s8b, which was the first one 
fabricated (a year ahead of the others), appears to be somewhat anomalous in terms of having $T_{\rmn{c}}$ and $G$ much closer 
to the specification. $G$ is controlled by the geometry of the silicon nitride membrane and it is known that phonon transport 
across thin film membranes at very low temperatures is a complex and poorly understood process, and may, for example, depend 
on factors such as the roughness of the membrane surface. This is a particularly important consideration for the ultra low NEP 
detectors needed for ground-based Cosmic Microwave Background experiments (where sensitivity is paramount over number of 
bolometers) and space-borne instruments of the future (where background power levels will be very low). From Table 2 it can be 
seen that expected phonon noise NEP, based on measured values of $G$ and $T_{\rmn{c}}$, is significantly higher than the 
requirement to ensure background limited performance at 850\,$\umu$m (Section 3.1.2; Table 1).

\begin{figure*} 
\centering 
\includegraphics[width=180mm]{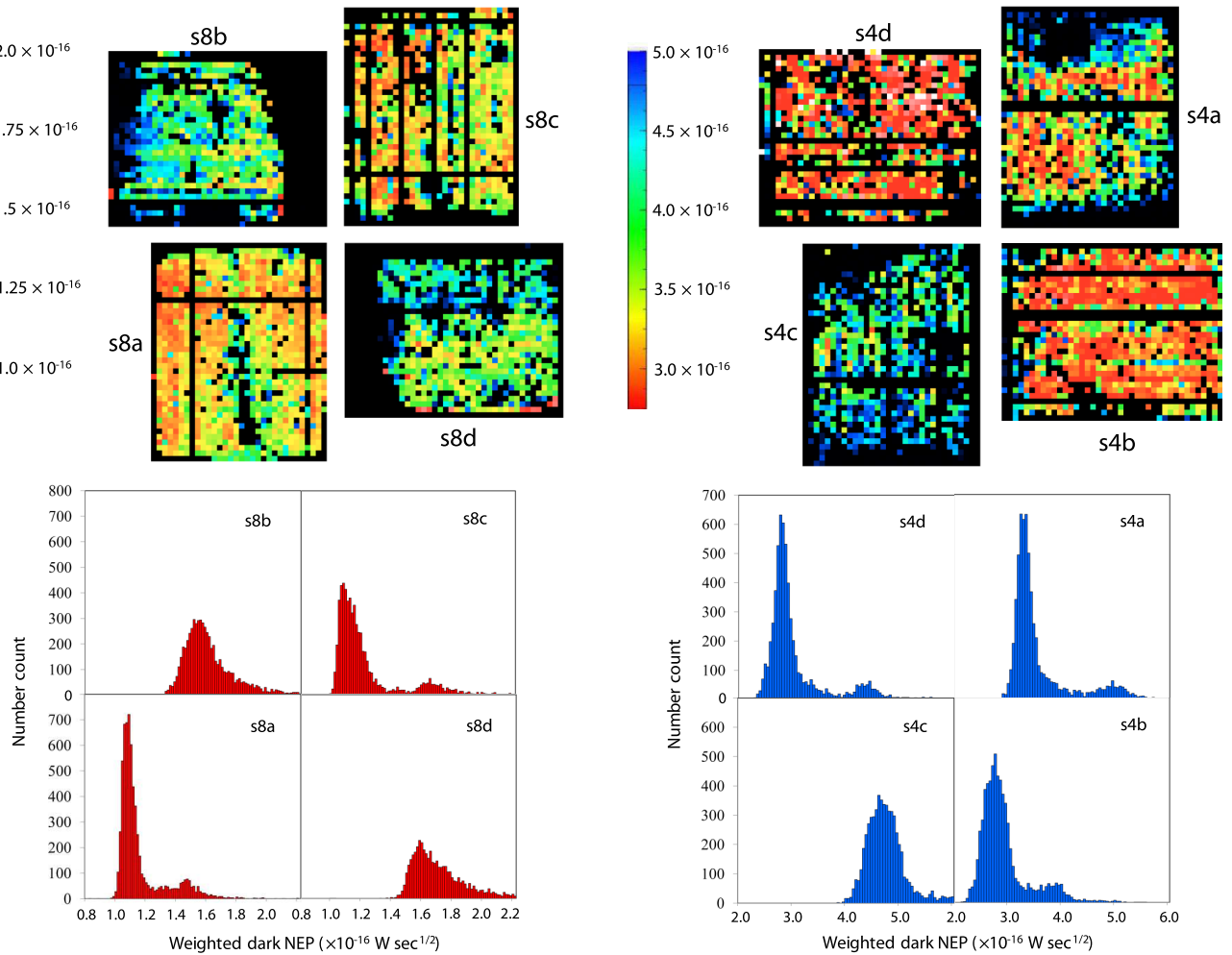} 

\caption{Top: Typical dark NEP images recorded with the shutter closed for each of the SCUBA-2 sub-arrays. These are derived 
from a ``dark noise'' measurements of $\sim$10\,sec at the start of each observation. The colour scale on each image 
represents the NEP in units of W\,s$^{1/2}$); Bottom: Dark NEP histograms for each sub-array, each on the same scale for 
ease of comparison. }

\label{fig:dark_neps} 
\end{figure*}

\subsubsection{Dark NEPs}

The noise equivalent power (NEP) is conventially defined as the signal power that gives a signal-to-noise ratio (SNR) of unity for 
an integration time of 0.5\,s. The dark NEP per bolometer is calculated from the ratio of the measured dark current noise 
(determined over a frequency range of 2--10\,Hz) to the responsivity (calculated from a ramp of the heater current; Section 
5.2)\footnote{The SCUBA-2 software calculates noise values for an integration time of 1\,s, and so the measured values in Table 2 
have been multiplied by $\sqrt{2}$ to allow for a comparison with the theoretical phonon NEP that assumes a post-detection 
bandwidth of 1\,Hz -- equivalent to an integration time of 0.5\,s, as given by the equation in Section 3.2.1.}. Taking simply the 
mean of the NEP of every bolometer per sub-array would be skewed by poorly performing detectors (as the distribution of values 
is non-Gaussian and so a weighted mean is used {\bf \footnote{The weighted NEP, ${\rmn{NEP}}_{\rmn{weight}} = \rmn{NEP}^{-2}\, 
{\rmn{NEP}}_{\rmn{mean}}$.}}. Since a dark-noise measurement is routinely carried out at the start of every astronomical 
observation a huge database of measurements now exists. The values given in Table 2 are a median value of 6,500 dark noise 
measurements taken between the period 2012 February and 2012 July. The measured dark NEP is typically 2--4 times higher than the 
expected phonon noise limited NEP. The high NEPs could be due to excess low frequency noise and/or lower-than-expected 
responsivities. There are several possible mechanisms to generate excess noise over the 2--10\,Hz range, including aliased noise 
from high frequency sources and effects due to magnetic flux trapped in the SQUIDs during cool-down (there is some evidence from 
the dark SQUID data that this could be a significant factor). The SCUBA-2 bolometers also exhibit excess noise at frequencies below 
1\,Hz with a typical ``1/$f$'' knee at around 0.7\,Hz. Although excess noise mechanisms are still under investigation, the source 
of 1/$f$ noise is believed to be largely intrinsic to the detector itself and not associated with the SQUIDs or readout circuit 
(based on measurements of the dark SQUID data). This fundamental limitation is the main reason why fast scanning modes had to be 
developed to move the signal frequencies beyond the 1/$f$ knee.

\subsubsection{Overall yield and stability}

Fig. 9 shows typical dark NEP ``images'' for each of the 8 sub-arrays and histograms of the NEP distribution. As can be seen, there are 
a number of non-functional bolometers. Some rows, columns and individual bolometers are faulty as a result of an issue during 
fabrication and show no response at all (e.g. a broken wire bond or non-functional SQ2 can knock out an entire column). Others are 
deliberately switched-off in a ``bad-bolometer'' mask, if, for example, they show sign of instability (e.g. an oscillating output). As 
well as the higher-than-expected $T_{\rmn{c}}$ and $G$, the variation of these properties across a given sub-array has performance and 
operational implications. There are some sub-arrays (e.g. s8b and s8d) that show distinct gradients or variations in NEP as a result of 
this. A single TES and heater bias (per sub-array) is insufficient to overcome these variations, resulting in regions of the sub-array 
where the bolometers are not biased into transition. Furthermore, other bolometers are less-than optimally biased in terms of minimum 
noise and maximum responsivity (i.e. minimum NEP). One possible way to smooth out the effects of the variation in $T_{\rmn{c}}$ is a 
novel technique called ``$T_{\rmn{c}}$ flattening''. By applying a higher SQ1 bias on selected rows for a short period in the MUX 
cycle, the SQ1 can be used as a secondary heater, thereby allowing rows of bolometers to be more optimally biased. With reference to 
Fig. 9 this would particularly benefit sub-arrays s8b, s8d and s4a. However, it is a limited technique in that it can only work on a 
row of bolometers and cannot correct for any $T_{\rmn{c}}$ variations across a row. At the time of writing this technique remains under 
investigation and is not currently implemented.

\vskip 1mm

The sub-array yields presented in Table 2 are the typical percentages of bolometers that contribute to an observation (these 
having been through a flat-fielding quality assurance test; Section 5.2). The average yield is about 70 per cent, which was 
the target goal at the start of the array design and fabrication process. Further quality assurance checks on the bolometers 
during the map-making process typically reject another 5 per cent of bolometers. The final map yields are therefore typically 
$\sim$\,65 per cent, corresponding to approximately 3700 (out of 5120 bolometers) operational in a focal plane. Whilst 
minimising the NEP at the same time as maximising the yield of a sub-array remains work in progress, the SCUBA-2 working 
bolometer counts are by far the highest of any submillimetre instrument. 

\vskip 1mm

The sub-array stability has significantly improved from the time when the instrument was first installed on the telescope. In 
the early commissioning phase, bolometers often became unstable during even modest slews of the telescope. This was attributed 
to pickup in the SQUID summing coil as the sub-arrays bisect the local magnetic field. Additional magnetic shielding in the 
instrument \citep{Craig2010} and enhancements to the array setup procedure improved the stability significantly, to such an 
extent that the majority of bolometers now remain stable during even the largest of scans. As a precaution, regular fast setups 
are still performed after a lengthy telescope slew. The SCUBA-2 bolometers can show occasional distinct jumps or steps in the time 
series data, most likely caused by cosmic ray events. The steps are now identified and corrected by an algorithm in the data 
reduction software (Section 8; \citealt{Chapin2012}). From repeated measurements it has been shown that the dark performance 
is usually stable and very repeatable, with less than 5 per cent variation in the dark NEP between successive measurements.

\begin{figure*} 
\centering 
\includegraphics[width=170mm]{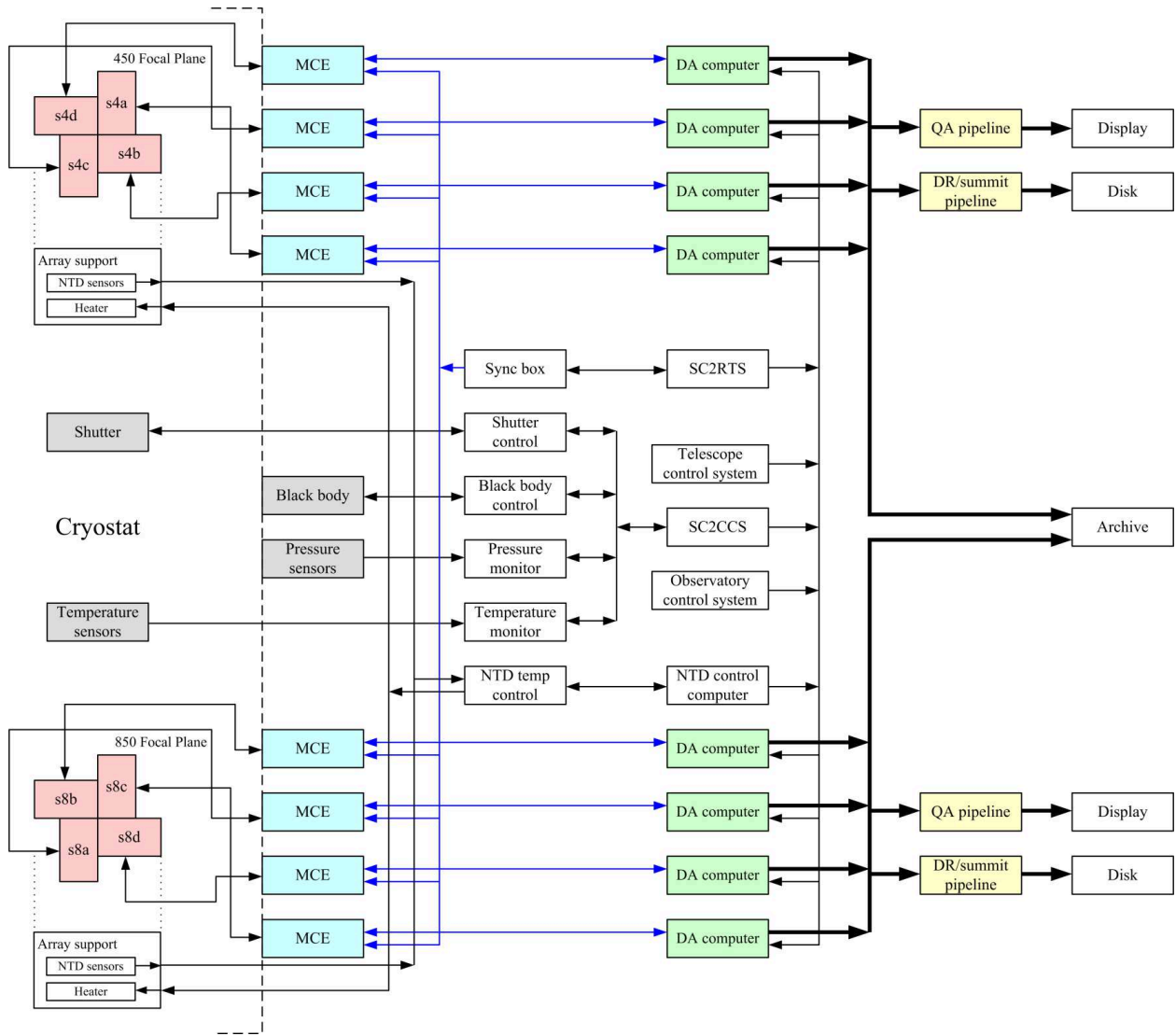} 

\caption{Schematic view of the data flow from the SCUBA-2 bolometer arrays showing the two pipelines running concurrently. The 
data from each sub-array is handled by a single data acquisition (DA) computer, which the pipelines read from to produce 
images. See Section 4 for an explanation of the acronyms.}

\label{fig:data_flow} 
\end{figure*}

\section{Signal and data processing}

The overall signal and data flow for SCUBA-2 are summarised in Fig. 10. This also includes monitoring of the instrument 
(temperatures and pressures) as well as temperature and mechanism control (SC2CCS). Each sub-array is read out using room 
temperature electronics (known as multi-channel electronics, or MCE) which in turn are each controlled by a data acquisition 
computer (DA). The data from the arrays transfers as frames at a rate of approximately 180\,Hz and are combined by the data 
reduction pipelines into images. The raw data and reduced images are stored on disk and transferred to the data archive 
centre. More details on the integration of SCUBA-2 into the JCMT observatory control system can be found in 
\citet{Walther2010}.

\subsection{Room temperature electronics and data acquisition}

The MCE is a self-contained crate that performs a number of functions. It sets the detector and heater bias, the bias (and 
feedback values as appropriate) for the three SQUID stages, controls the multiplexing rate and reads the DC-coupled signals 
from a 32 $\times$ 41 sub-array. In the standard data readout mode the MCE reports a low-pass filtered feedback value for 
every bolometer. There is one MCE crate per sub-array and the units are physically located on the outside of the main 
instrument. An address card in the MCE controls the time-division multiplexing by turning on one row of first stage SQUIDs at 
a time (see Fig. 7). Each bolometer is revisited at a rate of 13\,kHz (80\,$\umu$s) during the multiplexing, which far 
exceeds the bolometer response time. Separate readout cards are coupled to a set of 8 columns. As the current through a 
bolometer changes, as a result of power changes during an observation, a digital feedback servo (PID loop) is used to 
calculate the appropriate change to the feedback values sent to the SQ1 stage. Hence, these feedback vales represent a 
measurement of the optical power changes and are the nominal MCE outputs. The SQ1 signals of one column are summed in a coil 
coupled to one second-stage SQUID (Section 3.4.3). More information on the design and operation of the MCE can be found in 
\citet{Battistelli2008}.

\vskip 1mm

Each sub-array has a dedicated DA computer that sends commands to the MCE and receives data packets in return. The data 
acquisition software is based on a system running RTAI Linux. Data are packaged by the MCE into frames that consist of a 
house-keeping block followed by the data. The SCUBA-2 Real Time Sequencer (SC2RTS) coordinates and controls the tasks on each 
of the DA computers. The SC2RTS is a VME bus crate that takes commands from the main observatory RTS for coordinating 
instrument data-taking with the telescope actions. The sync box ensures that all sub-arrays clock out their data frames at 
exactly the same time. The data frames, together with house keeping information, are packaged by the DA computers into data 
files that are then subsequently passed to the data reduction pipelines. With SCUBA-2 operating in scan mode these data 
taking sequences can last up to 40\,min and contain many hundreds of thousands of frames. Since these datasets can be very 
large they are broken down into smaller sub-files typically written to disk every 30\,sec. The files are written to disk in 
Starlink NDF format \citep{Jenness2009} and contain header and house-keeping information. The 180\,Hz frame rate for SCUBA-2 
translates to a data rate of approximately 4\,MB s$^{-1}$ (raw, uncompressed data) at each wavelength. In terms of a 12\,hr 
observing night this is equivalent to typically 100\,GB of compressed data.

\subsection{Data reduction pipelines}

Data processing pipelines have been developed for SCUBA-2 using the established ORAC-DR pipeline infrastructure 
\citep{Cavanagh2008}. There are four pipelines running simultaneously at the telescope, two for each wavelength (see Fig. 10), 
which provide rapid feedback to observers on the quality of the data in real time. The ``quality assurance'' (QA) pipeline 
processes data for assessing the instrument performance and produces sensitivity estimates, flat-field updates and sub-array 
noise performance plots. The ``summit pipeline'' is designed to produce a quick-look map of the data. For the summit pipeline 
to run in real time it uses a curtailed version of the data reduction software described in Section 8. The pipeline can also 
be run in a highly-flexible and configurable off-line mode (``science pipeline''), making use of science data derived from the 
whole night (or multiple nights), and the optimised data reduction recipes available from the SMURF map-maker (Section 8.1). 
The data files are transferred to the JCMT Science Archive (JSA; Gaudet et al. 2008) at the Canadian Astronomy Data Centre 
(CADC) in Victoria \citep{Economou2011} within a few minutes of their appearance on disk. The primary aim of the JSA is to 
increase the productivity of the telescope by making science-ready data products available to the JCMT community. Hence, the 
data are reduced on a daily basis and fully processed images are made available to the project Principal Investigator within 
24\,hr.

\begin{figure*} 
\centering 
\includegraphics[width=110mm]{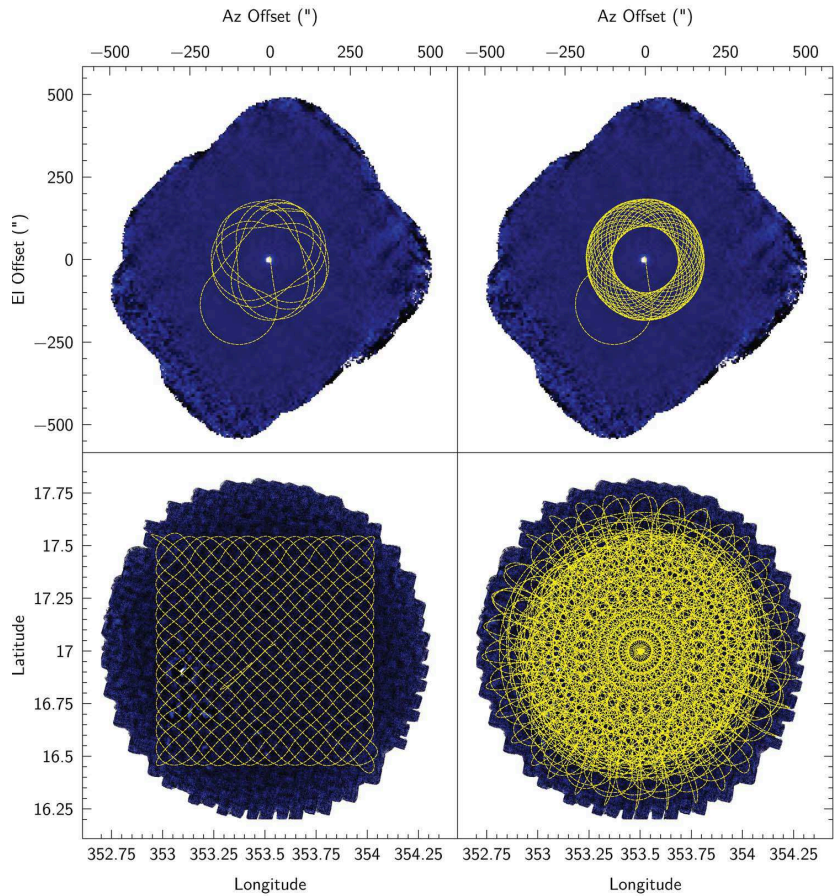} 

\caption{Telescope track in offsets of azimuth and elevation for the SCUBA-2 observing patterns. Top left: A single rotation of the 
{\sc daisy} pattern; Top right: Multiple rotations of the {\sc daisy} pattern for a typical map based on a 3\,arcmin demand 
diameter; Bottom left: A single {\sc pong} pattern; Bottom right: Multiple rotations of the {\sc pong} pattern for a typical map 
based on a 30\,arcmin demand diameter. {\bf The blue background represents the total covered area of sky during an observing 
pattern.}}

\label{fig:scan_modes} 
\end{figure*}

\section{Observing modes}

Since the major goal of SCUBA-2 is to conduct wide-field surveys of the sky the most efficient way to do this is to scan the 
telescope. To be able to recover large-scale structures in the presence of slowly-varying baselines (caused primarily 
by sky emission, extinction, and instrumental 1/$f$ noise) the scan pattern must modulate the sky both spatially and 
temporally in as many different ways as possible. Spatial modulation is achieved by scanning the same region at a number of 
different position angles to achieve cross-linking. Temporal modulation is incorporated by visiting the same region on 
different timescales. A number of scan patterns have been developed giving optimum coverage within the constraints of 
telescope motion \citep{Kackley2010}.

\subsection{Scan modes}

The telescope operates in a routine scanning mode for SCUBA-2 for which the type of scanning pattern adopted depends on the 
size of field to be observed. The scan pattern parameters (primarily the telescope speed and scan spacing) are chosen to ensure 
the effective integration times across the mapped region are as uniform as possible, as well as making it easy to define the 
shape of the region.

\subsubsection{Small-field observations}

For small fields, less than about the array footprint on the sky, constant speed ``{\sc daisy}'' scans are the preferred observing 
pattern\footnote{This was inspired by a similar mode used at the Green Bank Telescope}\footnote{Operationally, this is often referred 
to as a constant velocity {\sc daisy} scan}. In this mode the telescope moves in a pseudo-circular pattern that keeps the target 
coordinate on the arrays throughout the integration. The telescope is kept moving at a constant speed to maintain the astronomical 
signal at a constant frequency. The pattern on the sky is defined by two parameters: $R_{\rmn{0}}$, the radius of the requested map, 
and $R_{\rmn{T}}$, the turning radius. The optimisation of the {\sc daisy} observing mode involves identifying the parameters that 
provide a pattern that: (a) maximises the on-source integration time for a given elapsed time (minimising noise); and (b) gives uniform 
coverage within a 3\,arcmin diameter at the centre of the image. The {\sc daisy} scan pattern in Fig. 11 (top left and right) is 
optimised for the case in which $R_{\rmn{0}}$ and $R_{\rmn{T}}$ are both equal to 0.25 times the array footprint.

\vskip 1mm

The limitation of this mode is that the speed is constrained by the acceleration limit of the telescope (600 arcsec 
sec$^{-2}$ in true azimuth). When 1/cos(elevation) reaches $\sim$\,3 (elevation of $\sim$\,70$^\circ$) this acceleration 
limit is exceeded and the pattern tends to fail. Fig. 12 (top) shows the image plane and exposure time map for the standard 
{\sc daisy} pattern. Although the {\sc daisy} scan is designed for small and compact sources of order 3\,arcmin or less in 
diameter, there is significant exposure time in the map to more than double this size. The {\sc daisy} scan maximises the 
exposure time in the centre of the image. For example, an image in which the output map pixel sizes have been set to 2 and 
4\,arcsec (at 450 and 850\,$\umu$m, respectively), has an exposure time in the central 3\,arcmin region of $\sim$0.25 of the 
total elapsed time of an observation. Fig. 12 (top right) shows how the uniformity of the noise varies as a function of 
radius for a {\sc daisy} scan. Given that the noise level increases by 40 per cent at a radius of 3\,arcmin, this mode is 
useful for mapping point-like (unresolved) or compact objects of order 3--6\,arcmin in diameter and less. All calibration 
sources (Section 6) are observed with the {\sc daisy} scanning mode.

\begin{figure*} 
\centering 
\includegraphics[width=160mm]{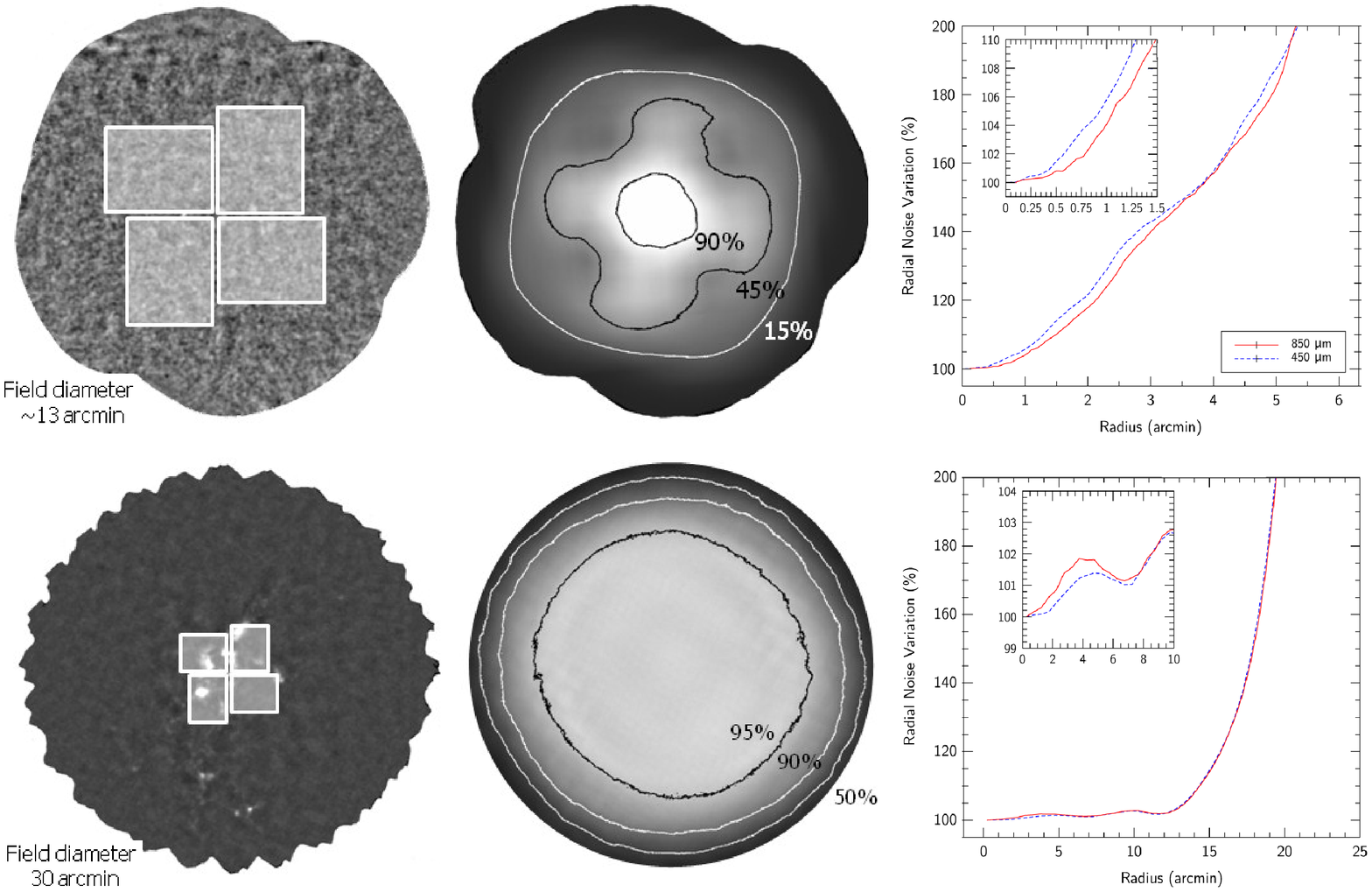} 

\caption{Left: Resultant images from a typical {\sc daisy} (top) and 30\,arcmin {\sc pong} (bottom) scan with the array 
footprint shown for scaling purposes; Middle: Exposure time images with contours at 90, 45 and 10\,per cent of the peak value 
for {\sc daisy} (top) and 95, 90 and 50\,per cent for {\sc pong} (bottom); Right: Radial noise profile in which the percentage 
increase in the RMS noise is plotted as a function of map radius, for {\sc daisy} (top) and 30\,arcmin {\sc pong} scan 
(bottom).}

\label{fig:exposure_time} 
\end{figure*}

\subsubsection{Large-field observations}

For fields just larger than the instrument field-of-view up to degree-sized scales, a map pattern called ``{\sc pong}'' is 
used as the scan mode\footnote{This is based on the ``bouncing billiard ball'' scan pattern developed for SHARC-II}. In this 
case the map area is defined to be square and the telescope tracks across the defined sky area, filling it in by ``bouncing'' 
off the walls of this area. A further innovation is to round-off the corners, making the transition at the walls curved and 
thereby keeping the telescope acceleration more uniform (``{\sc curvy pong}''). Once a pattern is completed the map is 
rotated and the pattern repeated at a new angle. This fulfils the criterion of cross-linking scans and providing as much 
spatial modulation as possible. Fig. 11 (bottom left and right) shows an example telescope track for a {\sc pong} map with a 
diameter of 30\,arcmin. The parameter space of the telescope speed, the spacing between successive rows of the basic pattern 
and the number of rotations have been optimised to give the most uniform coverage across the requested field. Fig. 12 
(bottom) shows the image plane and exposure time map for a 30\,arcmin diameter {\sc pong} pattern. The {\sc pong} scan 
maximises the field coverage and maintains even time uniformity. In this case output map pixel sizes of 2 and 4\,arcsec (at 
450 and 850\,$\umu$m, respectively) give an exposure time in the central 3\,arcmin region that is $\sim$\,0.014 of the 
elapsed time. Fig. 12 (bottom right) shows how the uniformity of the noise varies as a function of radius for a {\sc pong} 
scan. The noise remains uniform across the field, never increasing above 20 per cent relative to the centre of the map, out 
to the edge of the field.

\subsection{Flat-fielding}

The SCUBA-2 sub-arrays are flat-fielded using responsivity measurements derived from fast heater ramps. The bolometer signal 
current is determined from a series of different heater outputs consisting of a triangle wave of order a few pW (peak-to-peak) 
about a reference level. The inverse of a linear fit to the current as a function of heater power is the flat-field solution, with 
the responsivity (A\,W$^{-1}$) being the gradient. Bolometers are rejected that do not meet specific responsivity criteria (i.e. 
are deemed to be physically too low or high in value), if their response is non linear, or if the signal-to-noise ratio (SNR) of 
the measurement is poor. A flat-field measurement is performed at the start and end of every observation using a 5--10\,s 
repeating current ramp. The resulting flat-field is applied in the data reduction process for science maps (Section 8.1). The 
stability of the flat-field is usually excellent, with less than 1\,per cent variation in the number of bolometers meeting the 
acceptance criteria and less than 2\, per cent variation in mean responsivity on a sub-array over an entire night of 
observations.

\subsection{Pointing and focussing}

The telescope is accurately pointed and focussed using images derived from short {\sc daisy} scans of a bright, compact 
source. For pointing, a fitted centroid to the resultant image generates offsets from the nominal position. These are then 
passed back to the telescope control system to make adjustments in the azimuth/elevation position. For focus, an image is 
taken for each of 5 different offsets of the secondary mirror (in three-axes). A parabolic fit to the peak signal in each 
image generates an optimum focus offset which is passed to the secondary mirror controller.

\section{Calibration}

The calibration of ground-based submillimetre observations can be particularly problematic because of changes in the 
atmospheric opacity on short timescales \citep{Archibald2002}. The process of calibrating an observation requires two major 
steps. Firstly, the attenuation of the astronomical signal by the atmosphere is determined preferably along the line-of-sight. 
Secondly, astronomical images are calibrated by reference to a flux standard. The companion paper \citet{Dempsey2012} 
describes the calibration of SCUBA-2 data in more detail.

\subsection{Extinction correction}

The transmission of the atmosphere in the submillimetre is highly wavelength dependent (as shown in Fig. 3) and depends 
primarily on the level of PWV. At the JCMT weather conditions are categorised in terms of a ``weather band'' with a scale from 
1 to 5, with 1 being the driest and 5 the wettest. The weather band is derived from either direct measurements made at 
225\,GHz using a radiometer at the nearby CSO, or from a dedicated water vapour monitor (WVM) at the JCMT \citep{Wiedner2001}. 
The ``CSO tau'' measurement is derived from a fixed azimuth sky-dip (due south) and reports the zenith opacity every 15\,min. 
However, since the PWV can change on very short timescales at the JCMT, it is monitored at a faster rate using a separate WVM 
looking directly along the line-of-sight of the observation. The WVM estimates the level of PWV from the broadening of the 
183\,GHz water line in the atmosphere at intervals of 1.2\,s. Scaling the WVM measurement to a zenith opacity value shows 
a very close correlation to the ``CSO tau'', particularly during the most stable parts of the night (9\,pm until 3\,am) 
\citep{Dempsey2012}.

\vskip 1mm

Over the commissioning period the extinction relationships (at each SCUBA-2 waveband) with PWV and hence $\tau_{225}$ have 
been derived by analysing observations of sources of known flux density. The following relationships have been derived between 
the opacities at the SCUBA-2 wavebands and the 225\,GHz scaled measurements from the WVM:

\begin{equation}
\tau_{450} = 26.0 (\tau_{225} - 0.012);
\end{equation}

\begin{equation}
\tau_{850} = 4.6 (\tau_{225} - 0.0043).
\end{equation}

These relationships are subsequently used in the extinction correction stage during the process of making maps (Section 8.1).

\subsection{Flux calibration}

Primary calibration is taken from brightness temperature models of Mars \citep{Wright1976} and Uranus (Moreno, 2010\footnote{Moreno, R. 
''Neptune and Uranus brightness temperature tabulation'', ESA \emph{Herschel} Science Centre, ftp://ftp.sciops.esa.int/pub/hsc-calibration, 
2010}), and has been extended to include a number of compact ``secondary`` sources evenly spread over the sky. These secondary calibrators 
can take the form of late-type stars or compact H{\sc ii} regions. A flux conversion factor (FCF) is derived from the {\sc daisy} 
observation of a standard source and converts the raw bolometer signals into Janskys. The calibration of the bolometer heater (Section 
3.4.2) ensures that each sub-array in a focal plane reports the same optical power when observing an astronomical source and hence only a 
single FCF is needed at each waveband. The FCF depends on the photometry required for a particular source morphology and values are 
derived that are appropriate for both estimating the peak flux (usually applicable for an unresolved, point source) or the integrated flux 
(for an extended source). A database of secondary calibrators continues to be established to cover as much of the right ascension range as 
possible \citep{Dempsey2012}.

\section{On-sky performance}

\subsection{Typical observing sequence}

Each SCUBA-2 observation, based on either a {\sc daisy} or {\sc pong} observing pattern, follows an identical sequence. Once 
the telescope has been slewed to the appropriate source a fast array setup is carried out (Section 3.4.1). An observation then 
starts with a 10\,sec dark-noise measurement undertaken with the shutter closed. As the shutter opens to the sky, the power 
change is dynamically balanced by the heater tracking process (Section 3.4.2). Once the shutter is fully open and the power 
balance is stable, a flat-field measurement is carried out (Section 5.2). The heater carries out another small track at 
the end of the flat-field to compensate for any final sky power change. A science observation is then undertaken and typically 
lasts 30--40\,min, although pointing, focussing and calibration observations are much shorter (typically 5\,min). At the end 
of the observation there is another heater track before a final flat-field is carried out. Finally, the shutter closes and 
heater tracking restores the power balance to the dark value.

\subsection{On-sky sensitivity}

The sensitivity on the sky is represented by the noise equivalent flux density (NEFD) which is the flux density that produces 
a signal-to-noise of unity in 1\,s of integration time. At the shorter submillimetre wavelengths the NEFD is particularly 
heavily dependent on the weather conditions. The NEFD values are calculated in a similar way to the dark NEP (see Section 
3.5.2). A sky NEP value for each bolometer is calculated from the time series of the first sub-scan of an observation, and 
the responsivity as before from the flat-field measurement. The NEFD is then given by 
${\rmn{NEFD}}=({\rmn{NEP}}_{\rmn{sky}}{\rmn{FCF}}_{\lambda})/\eta$, where the FCF is the flux conversion factor determined 
from a flux calibrator (Section 6.2) and $\eta$ is the sky transmission. As in the case of the dark NEP a weighted average is 
used for the corresponding sky value. Fig. 13 shows how the NEFD varies as a function of sky transmission for both the 
SCUBA-2 wavebands. In terms of a direct bolometer-to-bolometer comparison, the SCUBA-2 values are 5--10\,per cent better than 
SCUBA at 450\,$\umu$m, and about the same at 850\,$\umu$m. The NEFD values in ``good'' observing conditions are typically 400 
and 90\,mJy sec$^{1/2}$ at 450 and 850\,$\umu$m, respectively, at least a factor of 2 worse than predicted based on a model of 
the instrument, telescope and Mauna Kea sky. A major contributor to these sensitivity figures is undoubtedly the 
higher-than-expected measured dark NEP (Section 3.5.2), although it is also possible that there are contributions from 
instrument and/or telescope that are not accounted for. This remains work under investigation.

\begin{figure} 
\centering 
\includegraphics[width=90mm]{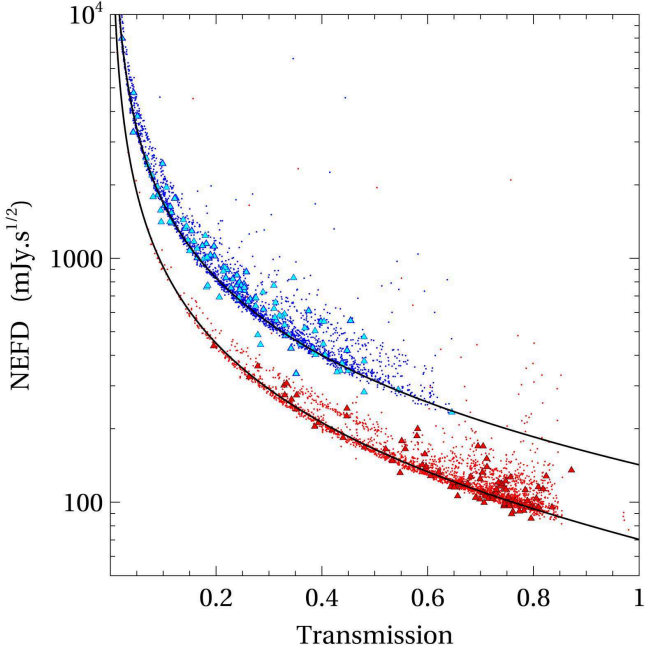} 

\caption{Measured NEFD for each SCUBA-2 waveband as a function of fractional sky transmission. The blue points are for 
450\,$\umu$m and the red points 850\,$\umu$m. The triangle symbols represent NEFD estimates taken directly from calibration 
observations using the measured RMS noise in the map and the FCF.}

\label{fig:nefd} 
\end{figure}

\subsection{Sensitivity limits and mapping speed}

The RMS noise in a map has been shown to integrate down as expected according to time$^{-1/2}$ as shown in Fig. 14 for 
a $\sim$7\,hr observation. In practical terms, a {\sc daisy} field of 3\,arcmin in diameter can reach a level of $\sim$1\,mJy at 
850\,$\umu$m in around 3\,hrs (in good conditions and including observing overheads), whilst for a 1\,deg diameter field a 
sensitivity limit of 6\,mJy can be obtained in about 7\,h. Table 3 lists a selection of detection limits for the SCUBA-2 
wavebands for various observing mode configurations.

\begin{figure} 
\centering
\includegraphics[width=85mm]{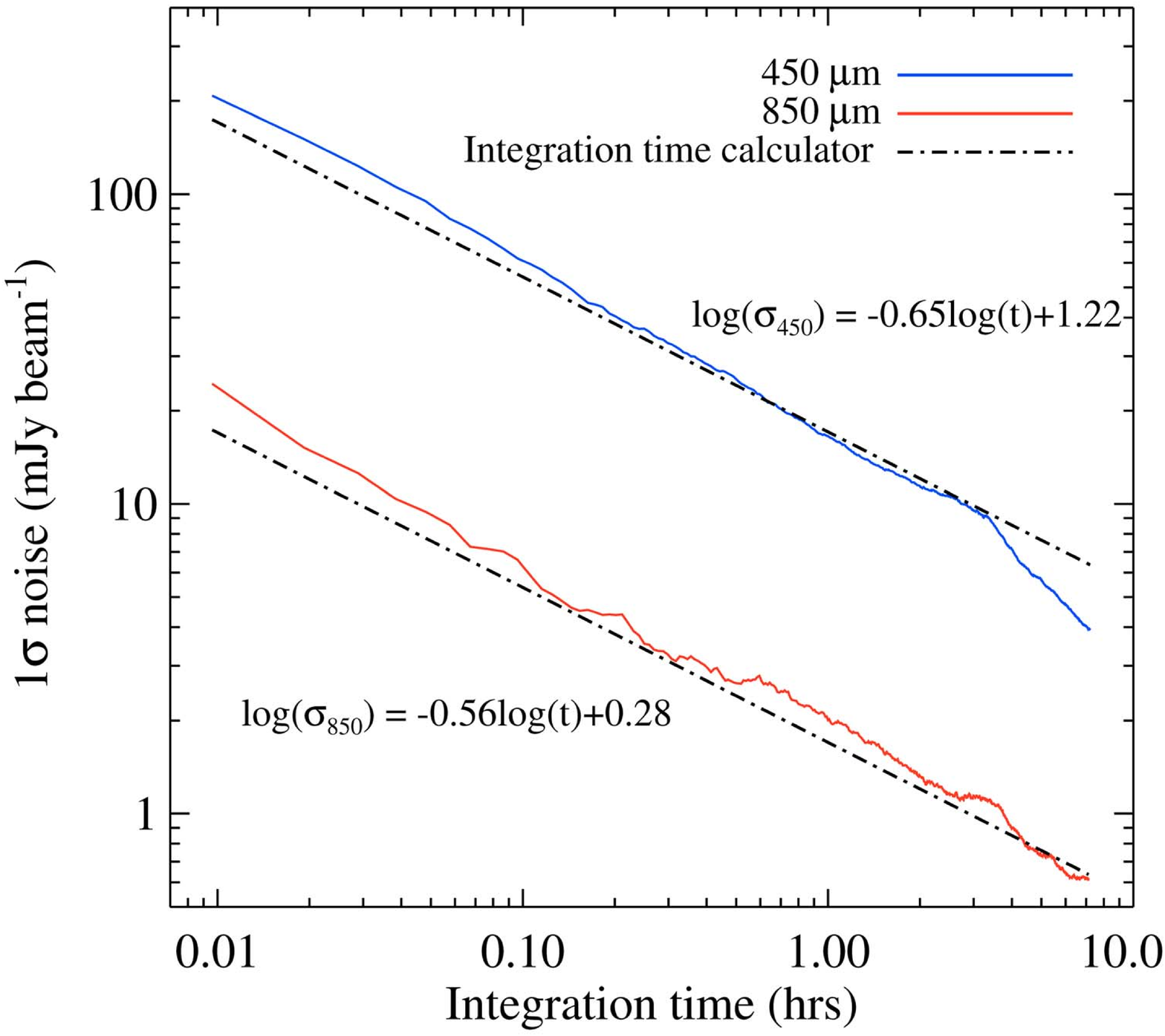} 

\caption{Measured RMS noise in a 3\,arcmin diameter {\sc daisy} image as a function of observing time for a $\sim$7\,h 
``blank-field'' observation taken in good observing conditions ($\tau_{225}$ $\sim$ 0.06). The equations for RMS noise as a 
function of observing time are as a result of a least-squares fit to the data (where the time, $t$, is in hours). The dashed 
lines represent the predicted RMS noise from the SCUBA-2 Integration Time Calculator based on a median opacity for the set of 
observations (see http://www.jach.hawaii.edu/jac-bin/propscuba2itc.pl). The observation represents a coadd of 11 separate 
observations taken over a period of several nights. As can be seen most evidently at 450\,$\umu$m the weather conditions 
improved significantly for some of the latter observations. Top: 450\,$\umu$m (blue); Bottom: 850\,$\umu$m (red). }

\label{fig:sensitivity_limits} 
\end{figure}

\vskip 1mm

Since the per-bolometer NEFDs are very similar to SCUBA, the SCUBA-2 mapping speed improvement is largely governed by the 
increase in detector count. Other factors include significantly lower observing overheads for the SCUBA-2 mapping modes than 
for the scan strategies used by SCUBA (e.g. no sky chopping). This results in mapping speed improvements of 100 and 150 times 
that of SCUBA for 450 and 850\,$\umu$m, respectively.

\vskip 1mm

\begin{table}
 \centering
  \caption{Detection limits in mJy for the SCUBA-2 observing modes. These have been calculated based on $\tau_{225}$ values of 
0.04 and 0.065 at 450 and 850\,$\umu$m, respectively, and assuming a source with average airmass of 1.2. The number associated 
with the {\sc pong} refers to the demand diameter of the map in arcsec.}
  \begin{tabular}{@{}llcc@{}}
  \hline

   Observing mode     &                        & 450\,$\umu$m            & 850\,$\umu$m     \\
                      &                        & (mJy)                   & (mJy)            \\
 \hline
   {\sc Daisy}        & (3\,$\sigma$,1\,h)    & 39                      & 5.6              \\
                      & (5\,$\sigma$,10\,h)   & 21                      & 2.9              \\
   {\sc Pong}900      & (3\,$\sigma$,1\,h)    & 85                      & 11.9             \\
                      & (5\,$\sigma$,10\,h)   & 44                      & 6.3             \\
   {\sc Pong}1800     & (3\,$\sigma$,1\,h)    & 166                     & 23               \\
                      & (5\,$\sigma$,10\,h)   & 87                      & 12.2             \\
   {\sc Pong}3600     & (3\,$\sigma$,1\,h)    & 361                     & 49              \\
                      & (5\,$\sigma$,10\,h)   & 189                     & 26               \\
   {\sc Pong}7200     & (3\,$\sigma$,1\,h)    & 732                     & 98              \\
                      & (5\,$\sigma$,10\,h)   & 384                     & 51               \\
\hline
\end{tabular}
\end{table}

\vskip 1mm

\subsection{Image quality}

Fig. 15 shows high signal-to-noise images of the beam shapes at 450 and 850\,$\umu$µm, based on, respectively a 54 and 80 
image mosaic of {\sc daisy} scans of Uranus (typical disk diameter of 3\,arcsec). The beams are fitted using two Gaussian 
components, namely a narrow main-beam and a wider secondary component. The main-beam widths (full-width at half-maximum), 
after de-convolving the Uranus disc, are 7.9 and 13.0\,arcsec at 450 and 850\,$\umu$m, respectively, whilst the secondary 
component has widths of 25 and 49\,arcsec. It is estimated that the main-beam widths are 6 and 2\,per cent higher than 
expected from a perfect optical system. The two component fit reveals that the main-beam has an amplitude of 94 and 98\,per 
cent at 450 and 850\,$\umu$m, respectively, which equates to integrated power levels of 60 and 75\,per cent (i.e. 40 and 
25\,per cent of the total power lies in the secondary component). The large ring visible in Fig. 15 is due to scalloping of 
the telescope panels since the focal length of the primary dish and panels are not exactly the same. This has an amplitude 
$\sim$\,0.1\,per cent of the peak at 450\,$\umu$m. Further details of the beam characterisation can be found in 
\citet{Dempsey2012}.

\begin{figure} 
\centering 
\includegraphics[width=85mm]{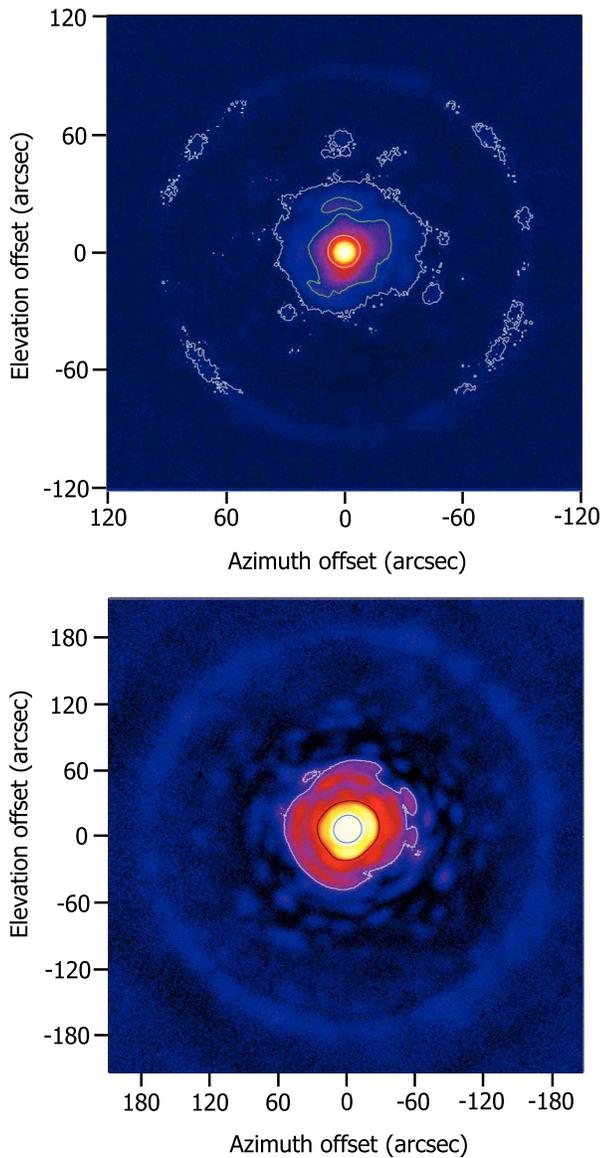} 

\caption{Measured beam using {\sc daisy} scanning of Uranus. Both plots have a log colour table to show the detail in the 
diffraction pattern. Top: 450\,$\umu$m with contours set at 0.1 (white), 1 (green) and 10\,per cent (white) of the peak 
amplitude, Bottom: 850\,$\umu$m with contours set at 0.1 (white), 1 (black) and 10\,per cent (blue).}

\label{fig:beam_patterns} 
\end{figure}

\vskip 1mm

To reconstruct maps to the highest possible degree of accuracy and image quality the relative position on the sky of each 
bolometer in the focal plane must also be determined. This is achieved by scanning every single bolometer in each focal plane 
across a bright source (such as Saturn or Mars), so that a map can be created from each bolometer. Since the position of the 
planet and telescope are known, the relative position of the bolometers can be determined. The results also demonstrate that 
there is very low field-distortion ($\sim$\,2\,per cent) across each focal plane.

\section{Data reduction and map-making}

SCUBA-2 data are reduced and images constructed using the Submillimetre User Reduction Facility (SMURF; see the companion 
paper Chapin et al. 2012), a software package written using the Starlink software environment \citep{Jenness2009}. By 
utilising SMURF within the data reduction pipeline fully-calibrated, publication quality images can be obtained.

\begin{figure} 
\centering 
\includegraphics[width=80mm]{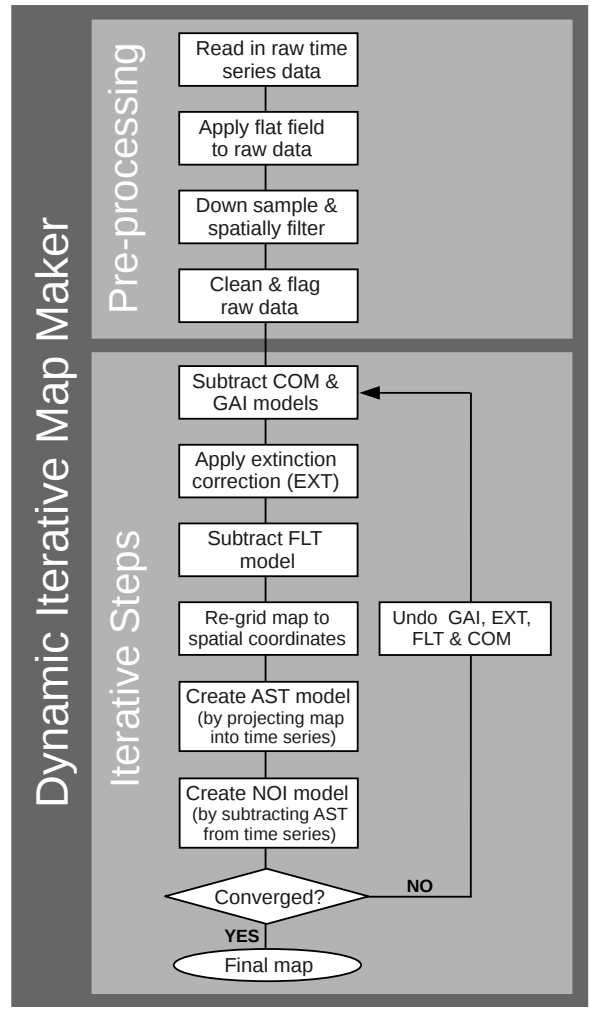} 

\caption{The SMURF map-making algorithm presented as a flowchart showing how the raw data is first pre-processed and then 
iteratively forms an output map through the application of a series of model components \citep{Chapin2012}.}

\label{fig:map_maker} 
\end{figure}

\subsection{Dynamic iterative map-maker}

The foundation of map-making within SMURF is an iterative technique that removes most of the correlated noise sources in 
parallel with a simplified map estimator. To accomplish this an overall model of the observed signal is constructed, breaking 
down the contributing components as appropriate. For example, the signal will have a time-varying component due to atmospheric 
extinction, a fixed astronomical source signature and various other sources of noise. The typical map-making algorithm is 
shown in the flowchart in Fig. 16. The initial step in the map-maker takes the individual sub-scans and combines the data into 
a contiguous time-series. Pre-processing applies the flat-field correction, re-samples the data at a rate that matches the 
requested output map pixel scale, and finally cleans the data by repairing spikes/DC steps and subtracting off a polynomial 
baseline from each bolometer.

\vskip 1mm

The iterative section then commences with estimating and removing a common-mode signal ({\sc com}), usually dominated by the 
atmosphere, and scaling it accordingly for each bolometer ({\sc gai}) so that a common calibration can be applied later for 
an entire sub-array. The {\sc com} model component is the average signal from all working bolometers on a sub-array at each 
time-step and flags bolometers as bad if their response does not resemble that from the majority of other bolometers. A 
time-dependent extinction correction factor ({\sc ext}) is then applied based on measurements from the WVM (Section 6.1). The 
data are subsequently Fourier transformed and a high-pass filter is applied to remove residual excess low-frequency noise 
({\sc flt}). The resulting cleaned and extinction-corrected data are re-gridded to produce an initial map estimate using 
nearest neighbour sampling. Since each map pixel will contain many bolometer samples the noise is significantly reduced 
compared to the raw time series data. The map is then projected back into the time domain, thus producing the {\sc ast} model 
containing signals that would be produced in each bolometer by the signal represented in the map. This model is then removed 
from the time-series data giving a residual signal from which the noise for each bolometer can be determined ({\sc noi}) with 
an associated value of $\chi^2$ used to monitor convergence. Since each signal component is slightly biased by signals from 
other components the entire process is iterated using a convergence tolerance. If the map has not changed from the previous 
iteration within this tolerance then the final output map is produced. If the map has changed, then the process is repeated.

\vskip 1mm

The map-making process is controlled by versatile configuration files that contain all the model settings and user-definable 
control parameters. For example, the high-pass filter cut-off is one parameter that can be easily adjusted. The convergence 
tolerance can also be bypassed by setting a fixed number of iterations. However, in reality, there are a small number of 
standard config files that are customised for use with different types of observations. More details can be found in 
\citet{Chapin2012}.

\section{Legacy surveys and early scientific results}

The key scientific driver for SCUBA-2 is the ability to carry out large-scale surveys of the submillimetre sky. Six 
``legacy-style'' survey programmes have been developed that are very broad-based, ranging from the studies of debris disks 
around nearby stars to galaxy populations and evolution in the early Universe. These surveys have been approved to run from 
2010 February 1 until 2014 September 30. In summary these surveys are\footnote{Further information on the survey programme can 
be found at: http://www.jach.hawaii.edu/jcmt/surveys}:

\begin{figure} 
\centering 
\includegraphics[width=85mm]{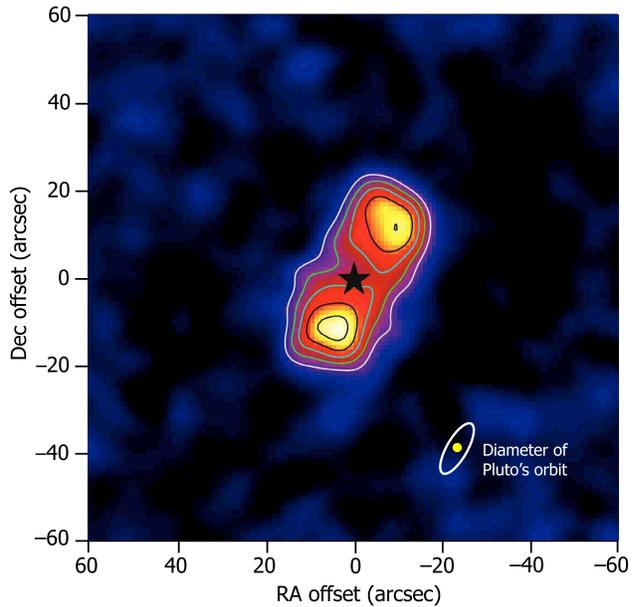} 

\caption{SCUBA-2 image of the debris disc around Fomalhaut at 850\,$\umu$m. Contours start at 3-$\sigma$ and increase in steps 
of 2-$\sigma$. The ``star'' symbol shows the position of Fomalhaut with respect to the disc. The diameter of Pluto's orbit in 
our Solar System is also shown, indicating that the disc may represent a Kuiper-Belt like structure around the star. Image 
provided courtesy of the SONS Legacy Survey team.}

\label{fig:fomalhaut} 
\end{figure}

\begin{itemize}

\item{Galactic Plane survey: 200+\,deg$^2$ to a 1-$\sigma$ depth of 15\,mJy at 850\,$\umu$m in 450\,h}

\item{Gould's Belt survey: Mapping of molecular clouds ($\sim$500\,deg$^2$) to 3\,mJy (850\,$\umu$m) in 412\,h}

\item{Debris Disk survey: Survey of 115 nearby stars to a depth of 1.4\,mJy (850\,$\umu$m) in 270\,h}

\item{Local Galaxy survey: Imaging of 150 nearby galaxies down to 1.6\,mJy (850\,$\umu$m) in 100\,h} 

\item{Cosmology survey (850\,$\umu$m): 10\,deg$^2$ (several fields) to 1.2\,mJy; (450\,$\umu$m): 0.25\,deg$^2$ to 
1.2\,mJy, for a total time of 1778\,h}

\item{SCUBA-2 ambitious sky survey: 1100\,deg$^2$ down to 30\,mJy at 850\,$\umu$m in 480\,h}

\end{itemize}

\vskip 1mm

Although the main strength of SCUBA-2 is in wide-field mapping, the camera can also image compact sources very quickly. Fig. 
17 is a short (2\,hour) 850\,$\umu$m observation of the famous debris disc surrounding the main sequence star Fomalhaut 
\citep{Holland2003} which extends to just under 1\,arcmin in length. Debris discs arise from collisions amongst 
planetetesimals in which the dusty residue spreads into a belt around the host star. Their study reveals much about the 
material left over after planet formation, the size of such systems compared with our own, the clearing out of comets that 
preceded the appearance of life on Earth, and even the detection of distant debris-perturbing planets (e.g. exo-Neptunes) that 
cannot be found by any other technique. The SCUBA-2 image took just one-fifth of the time of the previous SCUBA map to the 
same S/N level. Given that the per-bolometer NEFD values are very similar, the gain over SCUBA for compact and point-like 
source is largely due to not having to employ sky chopping to remove the atmosphere and ``jiggling'' of the seconday mirror to 
produce a Nyquist-sampled image. The disc is comparable to the size of the Kuiper Belt in our own Solar System and studying 
such discs therefore gives valuable insight into planetary system formation and evolution in our Galaxy.

\vskip 1mm

\begin{figure} 
\centering 
\includegraphics[width=80mm]{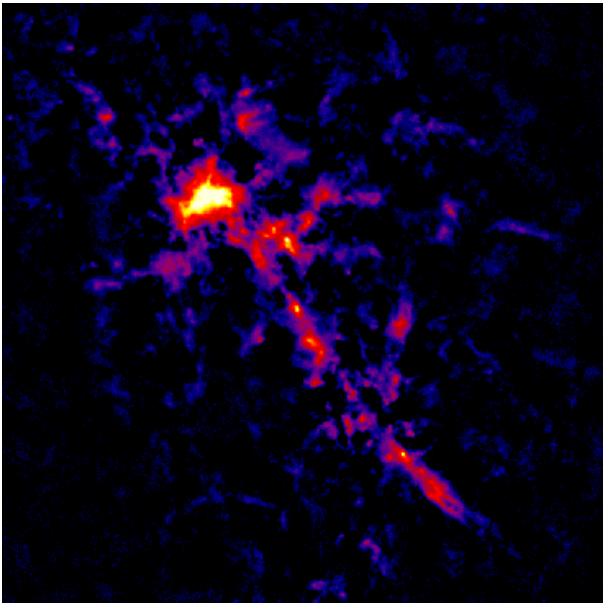} 

\caption{The high-mass star forming region W51 in Aquila as observed by SCUBA-2 at 850\,$\umu$m. The ridge of compact cores 
extending to the lower right in the figure runs parallel to the Galactic plane. The field size of this image is 
$\sim$\,1\,deg and the dynamic range is such that cores ranging in flux density from 40\,Jy to $<$\,20\,mJy are detected. 
Image provided courtesy of SCUBA-2 commissioning team.}

\label{fig:W51} 
\end{figure}

Wide-field imaging of sites of star formation in our own Galaxy is one of the key elements of several of the legacy surveys. A full 
understanding of the star formation process also requires an appreciation of how the rare, massive stars form and shape the 
evolutionary history of giant molecular clouds and subsequent star and planet formation. The early stages of high-mass star 
formation are not well understood, largely because they occur so fast and are consequently rare. A census of high-mass star 
formation throughout the Galaxy is possible with SCUBA-2. Fig. 18 shows a SCUBA-2 map at 850\,$\umu$m of the W51 star forming 
region, containing a ridge of massive star-forming cores runing parallel to the Galactic Plane. Studies such as this will show the 
rarest of evolutionary phases and allow an understanding of what defines the highest mass end of the stellar initial mass function. 
The sensitivity of SCUBA-2 equates to a mass sensitivity of 1\,M$_\odot$ at a distance of 3\,kpc and {bf 180\,M$_\odot$} at 
40\,kpc, sufficient to detect all the significant high-mass and cluster forming regions throughout the Galaxy.

\vskip 1mm

Another key area of the survey programme is to image the cold dust in nearby spiral galaxies. The bulk of star formation 
activity in nearby spirals is often missed by IR studies, since most of the dust mass resides in cold, extended, low-surface 
brightness discs, often far from the galactic nucleus. The studies so far have revealed that up to 90\,per cent of the total 
dust mass can be located within galactic discs. Dust temperatures are around 10--20\,K and so radiate strongly in the 
submillimetre region. Fig. 19 shows a \emph{Hubble Space Telecope (HST)} image of the famous ``Whirlpool galaxy'' M51 (and 
associated companion NGC\,5195) overlaid with SCUBA-2 colours (blue for 450\,$\umu$m; red for 850). SCUBA-2 clearly detects 
the nuclei of these two interacting galaxies and the fainter 850\,$\umu$m emission traces the optically-hidden dust lanes. 
Furthermore, the imaging power and spatial resolution achievable allows the study of regions of hot star formation in the 
outer arms of the spiral galaxy.

\begin{figure} 
\centering 
\includegraphics[width=90mm]{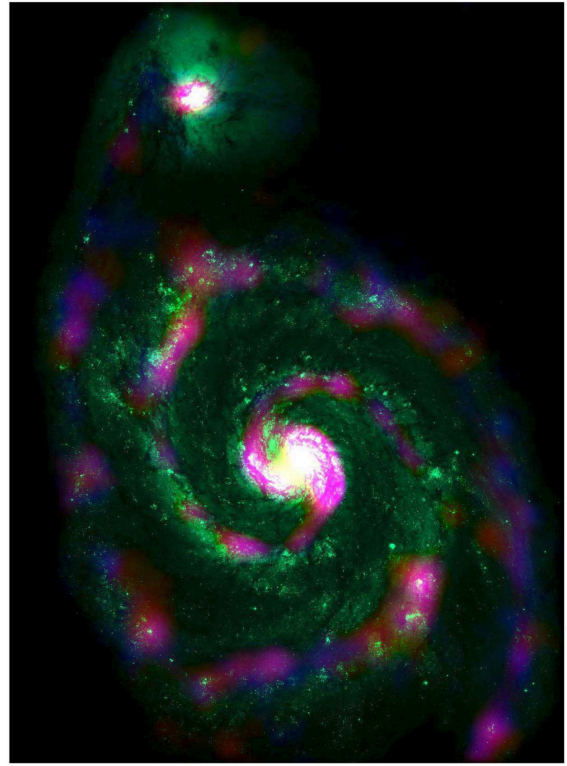} 

\caption{A composite image of the famous Whirlpool Galaxy with SCUBA-2 colours (blue for 450\,$\umu$m; red for 850) 
superimposed on a green-scale \emph{HST} image. SCUBA-2 traces star formation via the emission from cold dust in the outer 
most regions of galaxy. The SCUBA-2 image is provided courtesy of Todd MacKenzie, and the HST image is acredited to NASA, ESA, 
S. Beckwith (STScI) and the Hubble Heritage Team (STScI/AURA).}

\label{fig:m51} 
\end{figure}

\vskip 1mm

The final example of the versatility of SCUBA-2 is an observation of one of the most massive known cluster lenses, Abell 1689 
at $z$ = 0.18. Rich clusters are nature's telescopes that can be used to more efficiently study distant, star-forming 
galaxies. Figure 20 is a 850\,$\umu$m deep {\sc daisy} map of the Abell 1689 cluster field. The total field is approximately 
13\,arcmin in diameter and is known to contain over 50 lensed sources spanning a redshift range from 1--6. When SCUBA observed 
this field it detected 2 sources with SNR of greater than 4 and another 5 with tentative 3-$\sigma$ detections 
\citep{Knudsen2008}. SCUBA-2 imaged the field in a fraction of the time and detects 15 sources at greater than 5-$\sigma$ with 
many dozens at greater than 3-$\sigma$, confirming a mapping speed of over 100\,$\times$ SCUBA. SCUBA-2 is clearly a very 
powerful instrument for studying the distant Universe.

\section{Conclusions}

SCUBA-2 is the world's largest format camera for submillimetre astronomy. It represents a major step forward in submillimetre 
instrumentation in terms of the detector and array architecture, observing modes and dedicated data reduction pipelines. The 
new technologies developed for SCUBA-2 represent a major strategic investment on behalf of the JCMT and instrument funding 
agencies. The instrument has already shown incredible versatility with astronomy applications being very broad-based, ranging 
from the study of Solar System objects to probing galaxy formation in the early Universe. An imaging polarimeter 
\citep{Bastien2011} and Fourier transform spectrometer \citep{Gom2010} will also be available to allow the mapping of magnetic 
field lines and imaging medium-resolution spectroscopy, respectively.

\vskip 1mm

SCUBA-2 maps large-areas of sky 100--150 faster than SCUBA to the same depth, and such improved imaging power will allow the 
JCMT to exploit fully the periods of excellent weather on Mauna Kea. SCUBA-2 is currently undergoing a series of 6 unique legacy 
surveys for the JCMT community. These are highly complementary to the wider, but shallower, surveys undertaken by \emph{Herschel}, 
and are vital to fully exploit the capabilities of the new generation submillimetre interferometers and future facilities such 
as ALMA, CCAT and \emph{SPICA}.

\begin{figure} 
\centering 
\includegraphics[width=85mm]{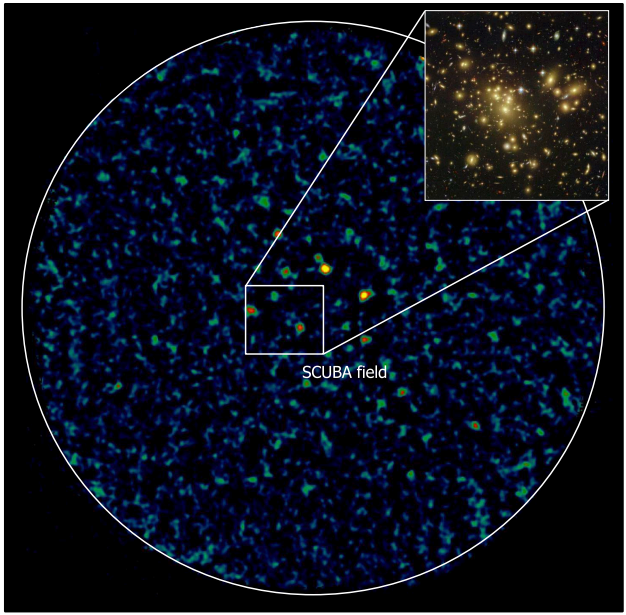} 

\caption{The massive lensing galaxy cluster Abell 1689 observed by SCUBA-2 at 850\,$\umu$m. The central inset shows the 
approximate region observed by SCUBA and the top right inset shows an \emph{HST} ACS image of the field. The new SCUBA-2 map 
detects 15 far-IR sources seen through the massive core of this cluster. The foreground mass amplifies the fluxes of the 
background source, enabling fainter sources to be detected, below the blank-field confusion limit of the JCMT. Image is 
provided courtesy of the SCUBA-2 Guaranteed-Time team.}

\label{fig:abell1689} 
\end{figure}

\section*{Acknowledgements}

The JCMT is operated by the Joint Astronomy Centre on behalf of the Science and Technology Facilities Council of the United 
Kingdom, the Netherlands Organisation for Pure Research, and the National Reserach Council of Canada. Additional funds for the 
construction of SCUBA-2 were provided by the Canada Foundation for Innovation.

\label{lastpage}


\begin{thebibliography}{99} 

\bibitem[\protect\citeauthoryear{Ade et al.}{2006}]{Ade2006} Ade P. A. R. Pisano G., Tucker C., Weaver S., 2006, in Zmuidzinas J., 
Holland W.S., Withington S., Duncan W. D., eds., Proc. SPIE Conf. Ser. Vol. 6275, Millimeter abd Submillimeter Detectors and 
Instrumentation for Astronomy III. SPIE, Bellingham, p. 62750U

\bibitem[\protect\citeauthoryear{Andr\'e et al.}{2010}]{Andre2010} Andr\'e P. et al., 2010, A\&A, 518, L102

\bibitem[\protect\citeauthoryear{Archibald et al.}{2002}]{Archibald2002} Archibald E. et al., 2002, MNRAS, 336, 1

\bibitem[\protect\citeauthoryear{Atad-Ettedgui et al.}{2006}]{Atad2006} Atad-Ettedgui E. et al., 2006, in Atad-Ettedgui E., 
Antebi J., Lemke D., eds., Proc. SPIE Conf. Ser. Vol. 6273, Optomechanical Technologies for Astronomy. SPIE, Bellingham, p. 62732H

\bibitem[\protect\citeauthoryear{Audley et al.}{2004}]{Audley2004} Audley M. D., Pisano G., Holland W. S., Duncan W. D., Parkes W., 
Ade P. A. R., 2004, in Zmuidzinas J., Holland W. S., Withington S., eds., Proc. SPIE Conf. Ser. Vol. 5498, Millimeter and Submillimeter 
Detectors for Astronomy II. SPIE Bellingham, p. 416

\bibitem[\protect\citeauthoryear{Bastien et al.}{2011}]{Bastien2011} Bastien P., et al., 2011, in Bastien, P., Manset N., Clemens D. 
P., St-Louis N., eds., ASP Conf. Ser. Vol. 449, Astronomical Polarimetry 2008: Science from Small to Large Telescopes. Astron. Soc. 
Pac., San Francisco, p. 68

\bibitem[\protect\citeauthoryear{Battistelli et al.}{2008}]{Battistelli2008} Battistelli E. S. et al., 2008, J. Low Temp Phys, 
151, 908

\bibitem[\protect\citeauthoryear{Bintley et al.}{2010}]{Bintley2010} Bintley D. et al., 2010, in Holland W. S., Zmuidzinas J., eds., 
Proc. SPIE Conf. Ser. Vol. 7741, Millimeter, Submillimeter, and Far-Infrared Detectors and Instrumentation for Astronomy V. SPIE, 
Bellingham, p. 774106

\bibitem[\protect\citeauthoryear{Bintley et al.}{2012a}]{Bintley2012a} Bintley D., Kuroda J. T., Starman E. G., Craig S. C., Holland 
W. S., 2012a, in Holland W. S., Zmuidzinas J., eds., Proc. SPIE Conf. Ser. Vol. 8452, Millimeter, Submillimeter, and Far-Infrared 
Detectors and Instrumentation for Astronomy VI. SPIE, Bellingham, p. 84523C

\bibitem[\protect\citeauthoryear{Bintley et al.}{2012b}]{Bintley2012b} Bintley D., et al., 2012b, in Holland W. S., Zmuidzinas J., 
eds., Proc. SPIE Conf. Ser. Vol. 8452, Millimeter, Submillimeter, and Far-Infrared Detectors and Instrumentation for Astronomy VI. 
SPIE, Bellingham, P. 845208

\bibitem[\protect\citeauthoryear{Blain et al.}{1999}]{Blain1999} Blain A., Smail I., Ivison R. J., Kneib, J.-P., 1999, MNRAS, 
302, 632

\bibitem[\protect\citeauthoryear{Cavanagh et al.}{2008}]{Cavanagh2008} Cavanagh B., Jenness T., Economou F., Currie M. J., 
2008, Astron. Nachr. / AN 329, 295

\bibitem[\protect\citeauthoryear{Chapin et al.}{2012}]{Chapin2012} Chapin E., Berry D. S., Gibb A. G., Jenness T., Scott D., 
Economou F., Holland, W. S., MNRAS in press

\bibitem[\protect\citeauthoryear{Craig et al.}{2010}]{Craig2010} Craig S. C. et al., 2010, in Holland W. S., Zmuidzinas J., eds., 
Proc. SPIE Conf. Ser. Vol. 7741, Millimeter, Submillimeter, and Far-Infrared Detectors and Instrumentation for Astronomy V. SPIE, 
Bellingham, p. 77411K

\bibitem[\protect\citeauthoryear{deKorte et al.}{2003}]{deKorte2003} deKorte P. A. J. et al., 2003, Rev. Sci. Instrum., 74, 3807

\bibitem[\protect\citeauthoryear{Dempsey et al.}{2012}]{Dempsey2012} Dempsey J. et al., MNRAS in press

\bibitem[\protect\citeauthoryear{Di Francesco et al.}{2007}]{DiFrancesco2007} Di Francesco J., Evans N.J. II., Caselli P., 
Myers P. C., Shirley Y., Aikawa Y., Tafalla M., 2007, in Reipurth B., Jewitt D., Keil K., eds., Protostars and Planets V, 
University of Arizona Press, Tucson, p. 17

\bibitem[\protect\citeauthoryear{Doriese et al.}{2007}]{Doriese2007} Doriese W. B. et al., 2007, Appl. Phys. Lett., 90, 193508 

\bibitem[\protect\citeauthoryear{Dowell et al.}{2002}]{Dowell2002} Dowell C. D. et al. 2003, Phillips T. G., Zmuidzinas J., 
eds., Proc. SPIE Conf. Ser. Vol. 4855, Millimeter and Submillimeter Detectors for Astronomy. SPIE, Bellingham, p. 73

\bibitem[\protect\citeauthoryear{Economou et al.}{2011}]{Economou2011} Economou F. et al., 2011, in Evans I. N., Accomazzi A., Mink 
D. J., Rots A. H., eds., ASP Conf. Ser. Vol. 442, Astronomical Data Analysis Software and Systems XVII. Astron. Soc. Pac., San 
Francisco, p. 203

\bibitem[\protect\citeauthoryear{Gao et al.}{2008}]{Gao2008} Gao X. et al., 2008, in Duncan W. D., Holland W. S., Withington S., 
Zmuidzinas J., eds., Proc. SPIE Conf. Ser. Vol. 7020, Millimeter and Submillimeter Detectors and Instrumentation for Astronomy IV. 
SPIE, Bellingham, p. 702025

\bibitem[\protect\citeauthoryear{Gaudet et al.}{2008}]{Gaudet2008} Gaudet, S., Dowler P., Goliath S., Redman R., 2008, in Bunclark 
P. S., Lewis J. R., eds., ASP Conf. Ser. Vol. 394, Astronomical Data Analysis Software and Systems XVII. Astron. Soc. Pac., San 
Francisco, p. 135

\bibitem[\protect\citeauthoryear{Glenn et al.}{1998}]{Glenn1998} Glenn J. et al., 1998, Phillips T. G., ed., Proc. SPIE Conf.  
Ser. Vol. 3357, Advanced Technoloy MMW, Radio, and Terahertz Telescopes. SPIE, Bellingham, p. 326G

\bibitem[\protect\citeauthoryear{Gom \& Naylor}{2010}]{Gom2010} Gom B., Naylor D. A., 2010, in Holland W. S., Zmuidzinas J., eds., 
Proc. SPIE Conf. Ser. Vol. 7741, Millimeter, Submillimeter, and Far-Infrared Detectors and Instrumentation for Astronomy V. SPIE, 
Bellingham, p. 77412E

\bibitem[\protect\citeauthoryear{Gostick et al.}{2004}]{Gostick2004} Gostick D., Montgomery D., McGregor H., Woodcraft A., Gannaway 
F., 2004, in Moorwood A. F. M., Iye M., eds., Proc. SPIE Conf. Ser. Vol. 5492, Ground-based Instrumentation for Astronomy. SPIE, 
Bellingham, p. 1743

\bibitem[\protect\citeauthoryear{Holland et al.}{1998}]{Holland1998} Holland W. S. et al., 1998, Nature, 392, 788

\bibitem[\protect\citeauthoryear{Holland et al.}{1999}]{Holland1999} Holland W. S. et al., 1999, MNRAS, 303, 659

\bibitem[\protect\citeauthoryear{Holland et al.}{2003}]{Holland2003} Holland W. S., et al., 2003, ApJ, 582, 1141

\bibitem[\protect\citeauthoryear{Hollister}{2009}]{Hollister2009} Hollister M. I., 2009, Ph.D Thesis, University of Edinburgh

\bibitem[\protect\citeauthoryear{Hollister et al.}{2008a}]{HollisterMagSh} Hollister M. I., McGregor H. Woodcraft A., Bintley D., 
MacIntosh M. J., Holland W. S., 2008, in Duncan W. D., Holland W. S., Withington S., Zmuidzinas J., eds., Proc. SPIE Conf. Ser. Vol. 
7020, Millimeter and Submillimeter Detectors and Instrumentation for Astronomy IV. SPIE, Bellingham, p. 702023

\bibitem[\protect\citeauthoryear{Hollister et al.}{2008b}]{HollisterDR} Hollister M. I., Woodcraft A., Holland W. S., Bintley D., 
2008, in Duncan W. D., Holland W. S., Withington S., Zmuidzinas J., eds., Proc. SPIE Conf. Ser. Vol. 7020, Millimeter and 
Submillimeter Detectors and Instrumentation for Astronomy IV. SPIE, Bellingham, p. 70200Y

\bibitem[\protect\citeauthoryear{Hughes et al.}{1998}]{Hughes1998} Hughes D. H. et al., 1998, Nature, 394, 241

\bibitem[\protect\citeauthoryear{Irwin}{1995}]{Irwin1995} Irwin K. D., 1995, Appl. Phys. Lett., 66, 1998

\bibitem[\protect\citeauthoryear{Irwin \& Hilton}{2005}]{Irwin2005} Irwin K. D., Hilton G. C., 2005, in Enss Chr., ed., 
Topics in Applied Physics, Vol. 99, Cryogenic Particle Detection. Springler-Verlag, Berlin Heidelberg, 63

\bibitem[\protect\citeauthoryear{Jenness et al.}{2009}]{Jenness2009} Jenness T., Berry D. S., Cavanagh B., Currie M. J., Draper P. 
W., Economou F., 2009, in Bohlender D. A, Durand D., Dowler P., eds., ASP Conf. Ser. Vol. 411, Astronomical Data Analysis and 
Systems XVIII. Astron. Soc. Pac., San Francisco, p. 418

\bibitem[\protect\citeauthoryear{Kackley et al.}{2010}]{Kackley2010} Kackley R. D., Scott D., Chapin E., Friberg P., 2010, in 
Radziwill N. M., Bridger A., eds., Proc. SPIE Conf. Ser. Vol. 7740, Software and Cyberinfrastructure for Astronomy. SPIE, 
Bellingham, p. 77401Z

\bibitem[\protect\citeauthoryear{Knudsen et al.}{2008}]{Knudsen2008} Knudsen, K. K., et al., 2008, MNRAS, 384, 1611

\bibitem[\protect\citeauthoryear{Kreysa et al.}{1998}]{Kreysa1998} Kreysa E. et al., 1998, in Phillips T. G., ed., Proc. SPIE 
Conf. Ser. Vol. 3357, Advanced Technoloy MMW, Radio, and Terahertz Telescopes. SPIE, Bellingham, p. 319

\bibitem[\protect\citeauthoryear{Mather}{1982}]{Mather1982} Mather J., 1982, Appl. Opt., 21, 1125

\bibitem[\protect\citeauthoryear{Motte et al.}{1998}]{Motte1998} Motte F., Andre P., Neri R., 1998, A\&A, 336, 150

\bibitem[\protect\citeauthoryear{Murphy et al.}{2011}]{Murphy2011} Murphy E. J., Chary R.-R., Dickinson M., Pope A., Frayer D. 
T., Lin L., 2011, ApJ, 732, 126

\bibitem[\protect\citeauthoryear{Siringo et al.}{2009}]{Siringo2009} Siringo G. et al., 2009, A\&A, 497, 945

\bibitem[\protect\citeauthoryear{Smail et al.}{1997}]{Smail1997} Smail I., Ivison R. J., Blain A., 1997, ApJ, 490, L5

\bibitem[\protect\citeauthoryear{Tucker \& Ade}{2006}]{Tucker2006} Tucker C., Ade P. A. R., 2006, in Zmuidzinas J., 
Holland W.S., Withington S., Duncan W. D., eds., Proc. SPIE Conf. Ser. Vol. 6275, Millimeter abd Submillimeter Detectors and 
Instrumentation for Astronomy III. SPIE, Bellingham, p. 62750T

\bibitem[\protect\citeauthoryear{Walther et al.}{2010}]{Walther2010} Walther C. A., Gao X., Kelly B. D., Kackley R. D., Jenness T., 
2010, in Radziwill N. M., Bridger A., eds., Proc. SPIE Conf. Ser. Vol. 7740, Software and Cyberinfrastructure for Astronomy. SPIE, 
Bellingham, p. 77400Y

\bibitem[\protect\citeauthoryear{Walton et al.}{2005}]{Walton2005} Walton A. J. et al., 2005, Journal of Nanoengineering and 
Nanosystems: Proc IMechE Part N, 219, 11

\bibitem[\protect\citeauthoryear{Wang et al.}{1996}]{Wang1996} Wang N. et al., 1996, Appl. Opt., 34, 6629

\bibitem[\protect\citeauthoryear{Ward-Thompson et al.}{2007}]{WardThompson2007} Ward-Thompson D., Andr\'e P., Crutcher R., 
Johnstone D., Onishi T., Wilson C., 2007, in Reipurth B., Jewitt D., Keil K., eds., Protostars and Planets V, University of 
Arizona Press, Tucson, p. 17

\bibitem[\protect\citeauthoryear{Wiedner et al.}{2001}]{Wiedner2001} Wiedner M., Hills R. E., Carlstrom J. E., Lay O. P., 
2001, ApJ, 533, 1036

\bibitem[\protect\citeauthoryear{Woodcraft et al.}{2009}]{Woodcraft2009} Woodcraft A. L., Hollister M. I., Bintley D., Gannaway F., 
Gostick D., Holland W. S., 2009, Cryogenics, 49, 504

\bibitem[\protect\citeauthoryear{Wright}{1976}]{Wright1976} Wright E. L., 1976, ApJ, 681, 415

\bibitem[\protect\citeauthoryear{Wyatt}{2008}]{Wyatt2008} Wyatt M. C., 2008, ARA\&A, 46, 339

\end{thebibliography}
\end{document}